\documentclass[useAMS,usenatbib]{mnras}
\usepackage{journals_macros}
\usepackage{graphicx,epstopdf}
\usepackage{natbib}
\usepackage{amsmath}
\usepackage{color}
\usepackage{balance}
\usepackage{amssymb}
\usepackage{multirow}
\usepackage{times}
\usepackage{xspace}
\usepackage{soul}
\usepackage{enumitem}
\usepackage[compatibility=false]{caption}
\usepackage{subcaption}
\usepackage{tabulary}
\usepackage[para]{threeparttable}
\usepackage{array,booktabs,longtable,tabularx}
\newcolumntype{L}{>{\raggedright\arraybackslash}X}
\usepackage{ltablex}
\usepackage{siunitx}
\usepackage[referable]{threeparttablex}
\renewlist{tablenotes}{enumerate}{1}

\makeatletter
\setlist[tablenotes]{label=\tnote{\alph*},ref=\alph*,itemsep=\z@,topsep=\z@skip,partopsep=\z@skip,parsep=\z@,itemindent=\z@,labelindent=\tabcolsep,labelsep=.2em,leftmargin=*,align=left,before={\footnotesize}}
\makeatother
\newcommand{\Mstardot}{M\ensuremath{_{\star}} / \mathrm{M}\ensuremath{_{\odot}}}

\newcommand{\fof}{FoF\xspace}
\newcommand{\fog}{FoG\xspace}

\newcommand{\com}{CoM\xspace}

\newcommand{\ssfr}{$\mbox{sSFR}$\xspace}

\newcommand{\hagn}{\mbox{{\sc \small Horizon-AGN}}}

\definecolor{Blue}{rgb}{0,0.08,0.65}
\definecolor{Red}{rgb}{0.65,0.08,0.05}
\definecolor{Green}{rgb}{0.35,0.45,0.25}
\definecolor{Orange}{rgb}{1.0,0.5,0.15}
\definecolor{Purple}{rgb}{0.5,0.0,0.5}

\usepackage{pifont}

\begin{document}
%
%
\title[Galaxy evolution in the metric of the Cosmic Web]{ 
       Galaxy evolution in the metric of the Cosmic Web}

  %
\author[K. Kraljic, S. Arnouts, C. Pichon et al.]{
\parbox[t]{\textwidth}{
\hskip -0.02cm
K. Kraljic$^{1}$\thanks{E-mail:katarina.kraljic@lam.fr},
         S. Arnouts$^{1}$, 
         C. Pichon$^{2,3}$,
         C.~Laigle$^{4}$,
         S. de la Torre$^{1}$,
         D. Vibert$^{1}$,
         C.~Cadiou$^{2}$,
         Y.~Dubois$^{2}$, 
         M.~Treyer$^{1}$,  
         C.~Schimd$^{1}$,
         S.~Codis$^{5}$,
         V.~de Lapparent$^{2}$,
         J.~Devriendt$^{4}$, 
         H.S.~Hwang$^{6}$,
         D.~Le Borgne$^{2}$,
         N.~Malavasi$^7$,
         B.~Milliard$^1$,
         M.~Musso$^{2,8}$,
         D.~Pogosyan$^{9}$,
          M.~Alpaslan$^{10}$,
         J.~Bland-Hawthorn$^{11}$,
         A. H.~Wright$^{12}$
         }
\vspace*{6pt} \\ 
$^{1}$ {Aix Marseille Univ, CNRS, LAM, Laboratoire d'Astrophysique de Marseille, Marseille, France} \\
$^{2}$ {Institut d'Astrophysique de Paris, UMR 7095 CNRS et Universit\'e Pierre et Marie Curie, 98bis Bd Arago, F-75014, Paris, France}\\
$^{3}$ {School of Physics, Korea Institute for Advanced Study (KIAS), 85 Hoegiro, Dongdaemun-gu, Seoul, 02455, Republic of Korea}\\
$^{4}$ {Sub-department of Astrophysics, University of Oxford, Keble Road, Oxford OX1 3RH}\\
$^{5}$ {Canadian Institute for Theoretical Astrophysics, University of Toronto, 60 St. George Street, Toronto, ON M5S 3H8, Canada}\\
$^{6}$ {Quantum Universe Center, Korea Institute for Advanced Study (KIAS), 85 Hoegiro, Dongdaemun-gu, Seoul 02455, Republic of Korea}\\
$^{7}$ {Department of Physics and Astronomy, Purdue University, 525 Northwestern Avenue, West Lafayette, IN 47907, USA}\\
$^{8}$ {Institut de physique th\'eorique, Universit\'e Paris Saclay and CEA, CNRS, 91191 Gif-sur-Yvette, France}\\
$^{9}$ {Department of Physics, University of Alberta, 412 Avadh Bhatia Physics Laboratory, Edmonton, Alberta, T6G 2J1, Canada}\\
$^{10}$ {NASA Ames Research Center, Moffett Field, Mountain View, CA 94035, United States}\\
$^{11}$ {Sydney Institute for Astronomy, School of Physics, University of Sydney, NSW 2006, Australia}\\
$^{12}$ {Argelander-Institut f\"ur Astronomie (AIfA), Universit\"at Bonn, Auf dem H\"ugel 71, D-53121 Bonn, Germany}
}
\date{submitted:  2017, accepted: }

\maketitle


\begin{abstract}
The role of the cosmic web in shaping galaxy properties is investigated in the GAMA spectroscopic survey in the redshift range $0.03 \leq z \leq 0.25$. The stellar mass, $u - r$  dust corrected colour and specific star formation rate (\ssfr) of galaxies are analysed as a function of their distances to the 3D cosmic web features, such as nodes, filaments and walls, as reconstructed by DisPerSE. Significant mass and type/colour gradients are found for the whole population, with more massive and/or passive galaxies being located closer to the filament and wall than their less massive and/or star-forming counterparts. Mass segregation persists among the star-forming population alone. The red fraction of galaxies increases when closing in on nodes, and on filaments regardless of the distance to nodes. Similarly, the star-forming population reddens (or lowers its \ssfr) at fixed mass when closing in on filament, implying that some quenching takes place. Comparable trends are also found in the state-of-the-art hydrodynamical simulation \hagn. These results suggest that on top of stellar mass and large-scale density, the traceless component of the tides from the anisotropic large-scale environment also shapes galactic properties. An extension of excursion theory accounting for filamentary tides provides a qualitative explanation in terms of anisotropic assembly bias: at a given mass, the accretion rate varies with the orientation and distance to filaments. It also explains the absence of type/colour gradients in the data on smaller, non-linear scales.

\end{abstract}


\section{Introduction}
\label{sec:intro}

\newcommand{\mean}[1]{
  \langle #1
  \rangle}
\newcommand{\tr}{\mathrm{tr}}
\newcommand{\nus}{\nu_{\!\S}}
\renewcommand{\S}{\mathcal{S}}
\newcommand{\sigmas}{\sigma_\S}
\newcommand{\dd}{\mathrm{d}}
\newcommand{\kk}{\mathbf{k}}
\newcommand{\Rs}{R_\S}

Within the $\Lambda$ cold dark matter ($\Lambda$CDM) cosmological paradigm, structures in the present-day Universe arise from hierarchical clustering, with smaller dark matter halos forming first and progressively merging into larger ones. Galaxies form by the cooling and condensation of baryons that settle in the centres of these halos \citep{WhiteRees1978}
and their spin is predicted to be correlated with that of the halo generated from the tidal field torques at the moment of proto-halo collapse \citep[tidal torque theory, TTT; e.g.][]{Peebles1969,Doroshkevich1970,EfstathiouJones1979,White1984}.
However, dark matter halos, and galaxies residing within them, are not isolated. They are part of a  larger-scale pattern,  dubbed the cosmic web \citep[][]{Joeveer1978,Bond1996}, arising  
from the anisotropic collapse of the initial fluctuations of the matter density field under the effect of gravity across cosmic time \citep{Zeldovich1970}. 
 
This web-like pattern, 
brought to light by systematic galaxy redshift surveys \citep[e.g.][]{deLapparent1986,geller1989,colless2001,Tegmark2004}, consists of large nearly-empty void regions surrounded by sheet-like walls framed by filaments which intersect at the location of clusters of galaxies. These are interpreted as the nodes, or high density peaks of the large-scale structure pattern, containing a large fraction of the dark matter mass \citep{BondMyers1996,Pogosyan1996}. The baryonic gas follows the gravitational potential gradients imposed by the dark matter distribution, then shocks and winds up around multi-stream, vorticity-rich filaments \citep{Codis2012,Laigle2015,Hahn2015}.
Filamentary flows, along specific directions dictated by the geometry of the cosmic web,  advect angular momentum into the newly formed low mass galaxies with  spins typically aligned with their neighbouring filaments \citep{Pichon2011,2013ApJ...769...74S}.
The next generation of  galaxies forms through mergers  as they drift  along these filaments towards the nodes of the cosmic web with a post merger spin preferentially perpendicular to the filaments, having converted they orbital momentum into spin \citep[e.g.][]{Aubert2004,Navarro2004,AragonCalvo2007a,Codis2012,Libeskind2012,Trowland2013,AragonCalvo2014,Dubois2014,Welker2015}.

Within the standard paradigm of hierarchical structure formation based on $\Lambda$CDM cosmology \citep{Blumenthal1984,Davis1985}, the imprint of the (\textit{past}) large-scale environment on galaxy properties is therefore to some degree expected via galaxy mass assembly history. 
Intrinsic properties, such as the mass of a galaxy (and  internal processes that are directly linked to its mass) are indeed shaped by its build-up process, which in turn is correlated with its \textit{present} environment. 
For instance, more massive galaxies are found to reside preferentially in denser environments \citep[e.g.][]{Dressler1980,PostmanGeller1984,Kauffmann2004,Baldry2006}.  
This mass-density relation can be explained through the biased mass function in the vicinity of the large-scale structure \citep[LSS;][]{Kaiser1984,Efstathiou1988} where   the enhanced density of the dark matter field  allows  the proto-halo to pass the critical threshold of collapse earlier \citep{Bond1991} resulting in an overabundance of massive halos in dense environments. 
However, what is still rightfully debated is whether the large-scale environment is also driving other observed trends such as morphology-density \citep[e.g.][]{Dressler1980,PostmanGeller1984,Dressler1997,Goto2003}, colour-density \citep[e.g.][]{Blanton2003,Baldry2006,Bamford2009} or star formation-density \citep[e.g.][]{Hashimoto1998,Lewis2002,Kauffmann2004} relations, and galactic `spin' properties, such as their angular momentum vector, their orientation, or chirality (trailing versus leading arms).

On the one hand, there are evidences that the cosmic web affects galaxy properties. 
Void galaxies are found to be less massive, bluer, and more compact than galaxies outside of voids \citep[e.g.][]{Rojas2004,Beygu2016}; galaxies infalling into clusters along filaments show signs of some physical mechanisms operating even before becoming part of these systems,
that galaxies in the isotropic infalling regions do not \citep[][]{Porter2008, Martinez2016}; 
\citet[][]{kleiner2017} find systematically higher HI fractions for massive galaxies ($M_{\star} > 10^{11} \mathrm{M}_{\odot}$) near filaments compared to the field population, interpreted as evidence for a  more efficient cold gas accretion from the intergalactic medium; \citet[][]{Kuutma2017} report an environmental transformation with a higher elliptical-to-spiral ratio when moving closer to  filaments, interpreted as an increase in the merging rate or the cut-off of gas supplies near and inside filaments \citep[see also][]{Aragoncalvo2016}; \cite{Chen2017} detect a strong correlation of galaxy properties, such as colour, stellar mass, age and size, with the distance to filaments and clusters, highlighting their role beyond the environmental density effect,
with red or high-mass galaxies and early-forming or large galaxies at fixed stellar mass having shorter distances to filaments and clusters than blue or low-mass and late-forming or small galaxies, and \cite{Tojeiro2017} interpret a steadily increasing stellar-to-halo mass ratio from voids to nodes for low mass halos, with the reversal of the trend at the high-mass end, found for central galaxies in the Galaxy And Mass Assembly survey \citep[]{Driver2009,Driver2011}, as an evidence for halo assembly bias being a function of geometric environment.
At higher redshift, a small but significant trend in the distribution of galaxy properties within filaments was reported in the spectroscopic survey VIPERS \citep[$z \simeq 0.7$;][]{Malavasi2017} and with photometric redshifts ($0.5< z< 0.9$) in the COSMOS field \citep[with a 2D analysis;][]{Laigle2017}. Both studies find important mass and type segregations, where the most massive or quiescent galaxies are closer to filaments than less massive or active galaxies, emphasising that large-scale cosmic flows play a role in shaping galaxy properties.

On the other hand, \cite{Alpaslan2015} find in the GAMA data
that the most important parameter driving galaxy properties is stellar mass as opposed to environment \citep[see also,][]{Robotham2013}. 
Similarly, while focusing on spiral galaxies alone, \cite{Alpaslan2016} do find variations in the star formation rate (SFR) distribution with large-scale environments, but they are identified as a secondary  effect.
Another quantity tracing different geometric environments that was found to vary is the luminosity function. However, while \cite{Guo2015} conclude that the filamentary environment may have a strong effect on the efficiency of galaxy formation \citep[see also][]{Benitez-Llambay2013},
\cite{Eardley2015} argue that there is no evidence of a direct influence of the cosmic web as these variations can be entirely driven by the underlying {\it local} density dependence.  
These discrepancies  are  partially expected:  the present state of galaxies must be impacted by the effect of the past environment, which  in turn does correlate with the present environment, if mildly so;  but these environmental effects must first be distinguished from mass driven effects which typically dominate.

The TTT, naturally connecting the large-scale distribution of matter and the angular momentum of galactic halos \citep[e.g.][]{Jones1979,Barnes1987,Heavens1988,Porciani2002a,Porciani2002b,Lee2004},
in its recently revisited, conditioned formulation \citep[][]{Codis2015} predicts the angular momentum distribution of the forming galaxies relative to the cosmic web, which tend to first have their angular momentum aligned with the filament's direction while the spin orientation of massive galaxies is preferentially in the perpendicular direction.
Despite the difficulty to model properly the halo-galaxy connection, due to the complexity, non-linearity and multi-scale character of the involved processes, modern cosmological hydrodynamic simulations confirm such a mass dependent angular momentum distribution of galaxies with respect to the cosmic web \citep[][]{Dubois2014,Welker2014,Welker2017}. 
On galactic scales, the dynamical influence of the cosmic web is therefore traced by the distribution of angular momentum and orientation of galaxies, when measured relative to their embedding large-scale environment. The impact of such environment on the spins of galaxies has only recently started to be observed \citep[confirming the spin alignment for spirals and preferred perpendicular orientation for ellipticals][but see also \citealp{Jones2010,Cervantes-Sodi2010,AndraeJahnke2011} for  contradictory results]{Trujillo2006,LeeErdogdu2007,Paz2008,Tempel2013,Tempel2013b,Pahwa2016}.  
What is less obvious is whether observed integrated scalar properties such as morphology or physical properties (star-formation rate, type, metallicity, which depend not only on the mass but also on the past and present gas accretion) are also impacted.

Theoretical considerations alone suggest that local density as a sole and unique parameter (and consequently any isotropic definition of the environment based on density alone) is not sufficient to account for the effect of gravity on galactic scale \citep[e.g.][]{Mo2010} and therefore capture the environmental diversity in which galaxies form and evolve:
one must also consider the relative past and present orientation of the tidal tensor with respect to directions pointing towards the  larger-scale structure principal axes.
At the simplest level, on large scales, gravity should be the dominant force.  Its net cumulative impact is encoded in the  tides operating on the host dark matter halo. Such tides may be decomposed into the trace of the tidal tensor, which equals the local density, and its traceless part, which applies distortion and rotation to the forming galaxy. The effect of the former on increasing scales has long been taken into account in standard galaxy formation scenarios \citep{Kaiser1984}, while the effect of the latter has only recently received full attention \citep[e.g.][]{Codis2015}. 
Beyond the above-discussed effect on  angular momentum, other galaxy's properties could in principle be influenced by the large-scale traceless part of the tidal field, which modifies the accretion history of a halo depending on its location within the cosmic web. For instance, the tidal shear near saddles along the filaments feeding massive halos is predicted to slow down the mass assembly of smaller halos in their vicinity \citep[][]{Hahn2009,Borzyszkowski2016,2016arXiv161103619C}.
 \cite{BondMyers1996} integrated the effect of ellipsoidal collapse (via the shear amplitude), which may partially delay galaxy formation, in the Extended Press-Schechter (EPS) theory. Yet, in that formulation, the geometry of the delay imposed by the specific relative orientation of tides imposed by the large-scale structure is not accounted for, because time delays are ensemble-averaged over all possible geometries of the LSS. 
The  anisotropy of the large-scale cosmic web -- voids, walls, filaments, and nodes  (which shape  and orient the tidal tensor beyond its trace)
should therefore be taken into account explicitly, as it impacts mass assembly.
Despite of the above-mentioned difficulty in properly describing the connection between galaxies and their host dark matter halos, this anisotropy should have direct observational signatures in the differential properties of galaxies with respect to the cosmic web at fixed mass and local density.
Quantifying these signatures is the topic of this paper.
Extending EPS  to account for the geometry of the tides beyond that encoded in the density of the field  is the topic  of the companion paper, \cite{biaspaper}.

This paper explores the impact of the cosmic web on galaxy properties in the GAMA survey, using the Discrete Persistent Structure Extractor code \citep[DisPerSE;][]{Sousbie2011a,Sousbie2011b} to characterise its 3D topological features, such as nodes, filaments and walls. 
GAMA is to date the best dataset for this kind of study, given its unique spectroscopic combination of depth, area, target density and high completeness, as well as  its broad  multi-wavelength coverage.
 Variations in stellar mass and colour, red fraction and star formation activity are investigated as a function of galaxy's distances to these three features. 
The rest of the paper is organised as follows. Section \ref{sec:data} summarises the data and describes the sample selection. The method used to reconstruct the cosmic web is presented in Section \ref{sec:methods}. 
Section~\ref{sec:mass_segregation} investigates the stellar mass and type/colour segregation and the star formation activity of galaxies within the cosmic web.
Section~\ref{sec:hagn} shows how these results compare to those obtained in the \hagn\,simulation~\citep{Dubois2014}.  
Section~\ref{sec:density} addresses the impact of the density on the measured gradients towards filaments and walls.
Results are discussed in Section~\ref{sec:discussion} jointly with predictions from \cite{biaspaper}. Finally, Section~\ref{sec:summary} concludes.
Additional details on the matching technique and the impact of the boundaries to the measured gradients are provided in Appendix~\ref{sec:appendix_matching} and~\ref{sec:appendix_boundaries}, respectively.
Appendix~\ref{sec:appendix_dtfe} investigates the effect of smoothing scale on the found gradients, 
Appendix~\ref{sec:appendix_hagn} briefly presents the horizon-AGN simulation, Appendix~\ref{sec:appendix_medians} provides tables of median gradients, and a short summary of predicted gradient misalignments is presented in Appendix~\ref{sec:gradient}. 

Throughout the study  a flat $\Lambda$CDM cosmology with H$_0 =$ 67.5 km s$^{-1}$ Mpc$^{-1}$, $\Omega_{M} = 0.31$ and $\Omega_{\Lambda} = 0.69$ is adopted \citep{PlanckXIII}. 
All statistical errors are computed by bootstrapping, such that the errors on a given statistical quantity correspond to the standard deviation of the distribution of that quantity re-computed in 100 random samples drawn from the parent sample with replacement. 
All magnitudes are quoted in the AB system, and by log we refer to the 10-based logarithm.

\section{Data and data products} 
\label{sec:data}

The following section describes the observational data and derived products, namely the galaxy and group catalogues, that have been used in this work.

\begin{figure*}
\includegraphics[width=\textwidth]{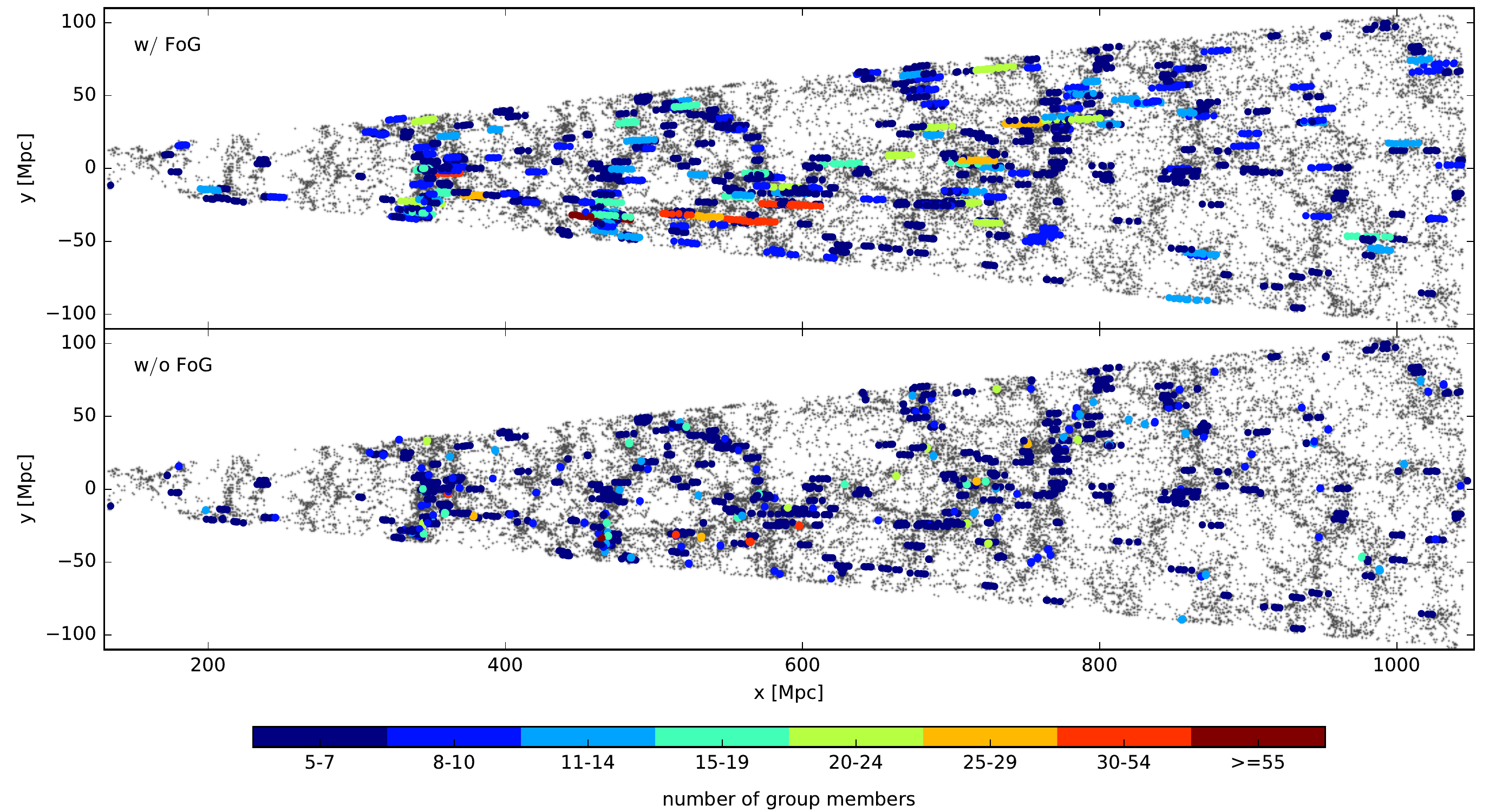} 
\caption{Spatial distribution of whole galaxy population with $m_r < 19.8$ in the GAMA field G12 in the redshift range $0.03 \leq z \leq 0.25$ (grey points).   
Overplotted are galaxy group members, colour coded by the size of their group. Only groups having five or more members are shown. The top and bottom panels illustrate the galaxy group members before and after correcting for the \fog effect, respectively. 
}
\label{Fig:data_groups}
\end{figure*}

\subsection{Galaxy catalogue}   
\label{subsec:galaxies}

The analysis is based on the GAMA survey\footnote{\href{http://www.gama-survey.org/}{http://www.gama-survey.org/}} \citep{Driver2009,Driver2011,Hopkins2013,Liske2015}, a joint European-Australian project combining  multi-wavelength photometry (UV to far-IR)  from ground and space-based facilities and spectroscopy obtained at the Anglo-Australian Telescope (AAT, NSW, Australia) using the AAOmega spectrograph. 
GAMA provides spectra for galaxies across five regions, but  this work  only considers  the three equatorial fields G9, G12 and G15 covering a total area of 180 $\deg^2$ ($12 \times 5 \deg^2$ each), for which
the spectroscopic completeness is $> 98$\% down to a $r$-band apparent magnitude $m_r = 19.8$.
The reader is referred to \citet[][]{Wright2016} for a complete description of the spectro-photometric catalogue constructed using 
the LAMBDAR\footnote{Lambda  Adaptive  Multi-Band  Deblending  Algorithm  in  R} code that was applied to  the  21-band
photometric  dataset  from  the GAMA Panchromatic Data Release \citep{Driver2016}, containing imaging spanning the far-UV
to the far-IR. 
 
The physical parameters for the galaxy sample such as the  absolute magnitudes, extinction corrected rest-frame colours, stellar masses and specific star formation rate (\ssfr)
are derived using a grid of model spectral energy distributions \citep[SED;][]{BruzualCharlot2003} and the SED fitting code LePHARE\footnote{\href{http://cesam.lam.fr/lephare/lephare.html}{http://cesam.lam.fr/lephare/lephare.html}}\citep{Arnouts1999,Ilbert2006}. The details  used to derive these physical parameters are given in the companion paper Treyer et al. (in prep.). 

The classification between the active (star-forming) and passive (quiescent) populations is based on a simple colour cut at $u - r = 1.8$ in the rest-frame extinction corrected $u - r$ vs $r$ diagram that 
  is used to separate the two populations. This colour cut is consistent with a cut in \ssfr{} at 10$^{-10.8}$ yr$^{-1}$ (see Treyer et al. in prep.). Hence, in what follows, the terms red (blue) and quiescent (star-forming) will be used interchangeably. 

The analysis is restricted to the redshift range $0.03 \leq z \leq 0.25$, totalling $97072$ galaxies. This is motivated by the high galaxy sampling required to reliably reconstruct 
 the cosmic web. Beyond  $z\sim 0.25$, the galaxy number density drops substantially (to $2 \times 10^{-3}$ Mpc$^{-3}$ from $8 \times 10^{-3}$ Mpc$^{-3}$ at $z \leq 0.25$, on average), while below $z\sim 0.03$, the small volume does not allow us to explore the large scales of the cosmic web.

The stellar mass completeness limits are defined for the passive and active galaxies as the mass above which 90\% of galaxies of a given type (blue/red) reside at a given redshift $z \pm 0.004$.
This translates into mass completeness limits of $\log \, (\Mstardot) = 9.92$ and $\log \, (\Mstardot) = 10.46$ for the blue and red populations at $z \leq 0.25$, respectively. 

\subsection{Group catalogue}    
\label{subsec:fof}

Since the three-dimensional distribution of galaxies relies on the redshift-based measures of distances, it is affected by their peculiar velocities.  
In order to optimise the cosmic web reconstruction, one needs to take into account these redshift-space distortions. On large scales, these arise from the coherent motion of galaxies accompanying the growth of structure, causing its flattening along the line-of-sight, the so-called Kaiser effect \citep{Kaiser1987}. On small scales, the co-called Fingers of God \cite[FOG;][]{Jackson1972,Tully1978} effect, induced by the random motions of galaxies within virialized halos (groups and clusters) causes the apparent elongation of structures in redshift space, clearly visible in the galaxy distribution in the GAMA survey (Figure~\ref{Fig:data_groups}, top panel). 
While the Kaiser effect tends to enhance the cosmic web by increasing the contrast of filaments and walls \cite[e.g.][]{SubbaRao2008,Shi2016}, the \fog effect may lead to the identification of spurious filaments. Because the impact of the Kaiser effect is expected to be much less significant than that of the \fog \cite[e.g.][]{SubbaRao2008,Kuutma2017}, for the purposes of this work, its correction is not attempted and the focus is on the compression of the \fog alone. 
To do so, the galaxy groups are first constructed with a use of an anisotropic Friends-of-Friends (\fof) algorithm operating on the projected perpendicular and parallel separations of galaxies, 
that was calibrated and tested using the publicly available GAMA mock catalogues of \cite{Robotham2011} \citep[see also][for details of the mock catalogues construction]{Merson2013}. Details on the construction of the group catalogue and related analysis of group properties can be found in the companion paper Treyer et al. (in prep.). 
Next, the centre of each group is identified following \cite{Robotham2011} \citep[see also][for a different implementation]{Eke2004}. The method is based on an iterative approach: first, the centre of mass of the group (\com) is computed; next its projected distance from the \com  is found iteratively for each galaxy in the group by rejecting the most distant galaxy. This process stops when only two galaxies remain and the most massive galaxy is then identified as the centre of the group.  
The advantage of this method, as shown in \cite{Robotham2011}, is that the iteratively defined centre is less affected by interlopers than luminosity-weighted centre or the central identified as the most luminous group galaxy. The groups are then compressed radially so that the dispersions in transverse and radial directions are equal, making the galaxies in the groups isotropically distributed about their centres \citep[see e.g.][]{Tegmark2004}. 
In practice, since the elongated \fog effect affects mostly the largest groups, only groups with more than six members are compressed.
Note that the precise correction of the \fog effect is not sought. What is needed for the purpose of this work is the elimination of these elongated structures that could be misidentified as filaments.

Figure~\ref{Fig:data_groups} displays the whole galaxy population and the identified \fof groups (coloured by their richness) in the GAMA field G12. The top and bottom panels show the groups before and after  correcting for the \fog effect. For the sake of clarity, only groups having at least five members are shown. The visual inspection reveals that most of the groups are located within dense regions, often at the intersection of the 
apparently filamentary structures.

\section{The cosmic web extraction}
\label{sec:methods}
With the objective of exploring the impact of the LSS on the evolution of galaxy properties, one first needs to properly describe the main components of the cosmic web, namely the high density peaks (nodes) which are connected by filaments, framing the sheet-like walls, themselves surrounding the void regions.
Among the various methods developed over the years, two broad classes can be identified. 
One uses the geometrical information contained in the {\it local} gradient and the Hessian 
of the density or potential field \citep[e.g.][]{Novikov2006, AragonCalvo2007b,AragonCalvo2007a,Hahn2007a,Hahn2007b,Sousbie2008a,Sousbie2008b,Forero-Romero2009,Bond2010b,Bond2010a}, while the second exploits the topology and connectivity of the density field  by using the  watershed transform \citep[]{AragonCalvo2010b} or  Morse theory \citep[e.g.][]{Colombi2000, Sousbie2008a,Sousbie2011a}. 
The theory for the former can be built in some details \citep[see e.g.][]{Pogosyan2009}, shedding some light on  physical interpretation,
while the latter avoids shortcomings of a second order Taylor expansion of the field and provides a natural metric in which to compute distances to filaments.
Within these broad categories, some algorithms deal with discrete data sets, while others 
require that the density field must be first estimated (possibly on multiple scales).
An exhaustive description of several cosmic web extraction techniques and a comparison of their classification patterns as measured in simulations are presented in \cite{Libeskind2017}. While this paper found some differences between the various algorithms, which should 
in principle be accounted for as modelling errors in the present work, these differences remain small on the scales considered. 

\subsection{Cosmic web with DisPerSE}
\label{subsec:disperse}
\begin{figure*}
\includegraphics[width=\textwidth]{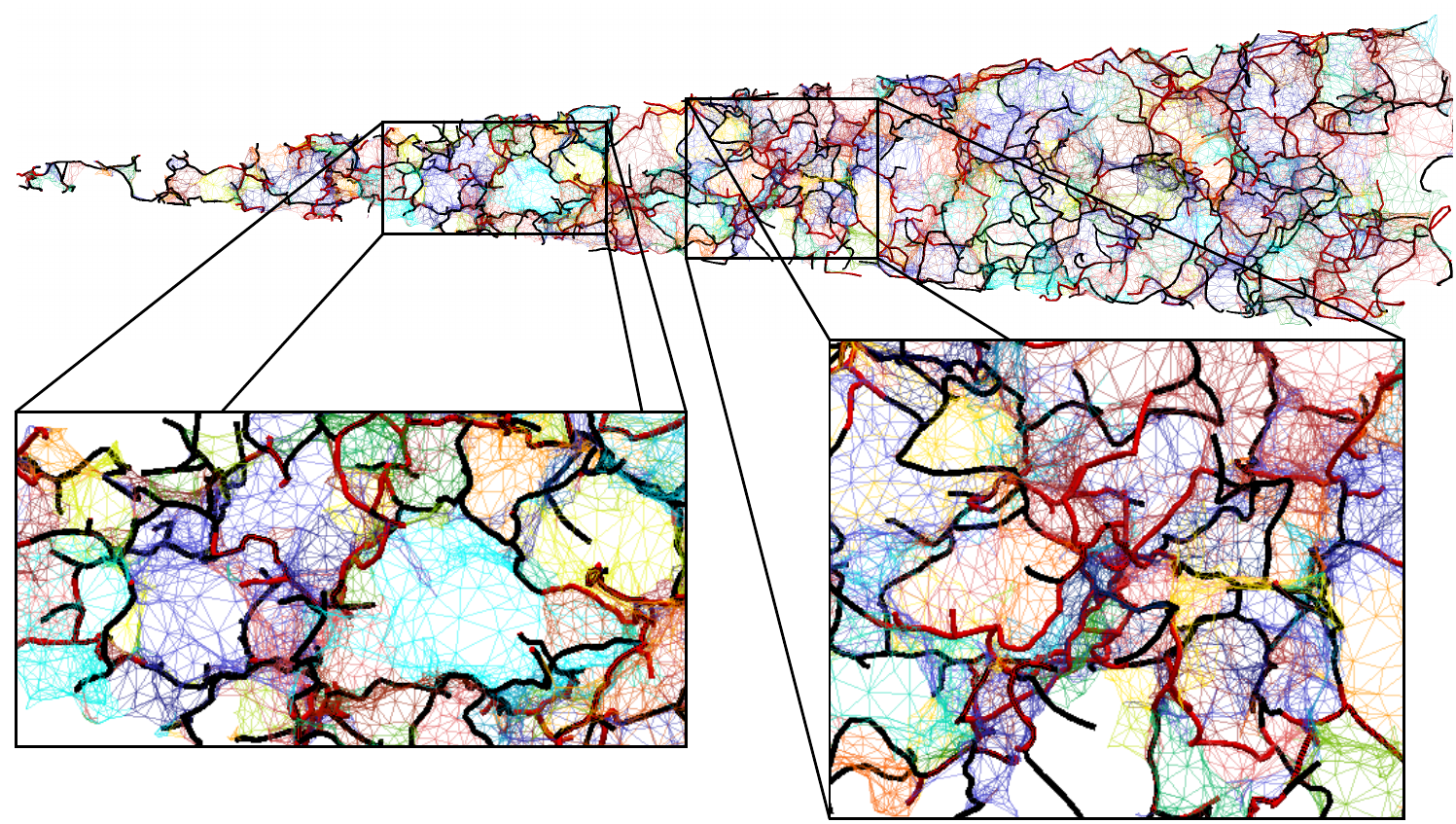}
\caption{
Illustration of the walls and filaments in the G12 field.
For the sake of clarity and for the illustrative purposes, only the cosmic web features detected above a persistence threshold of 5$\sigma$ are shown. Filaments are coloured in black, with the most persistent ones ($> 6\sigma$) plotted in red, while walls are colour coded randomly.     
Note how DisPerSE is capable of recovering the important features of the underlying cosmic field by identifying its (topologically) most-robust features. 
In particular, it extracts filaments as a set of connected segments, which outskirt walls, themselves circumventing voids.}
\label{Fig:walls}
\end{figure*}
\begin{figure*}
\includegraphics[width=\textwidth]{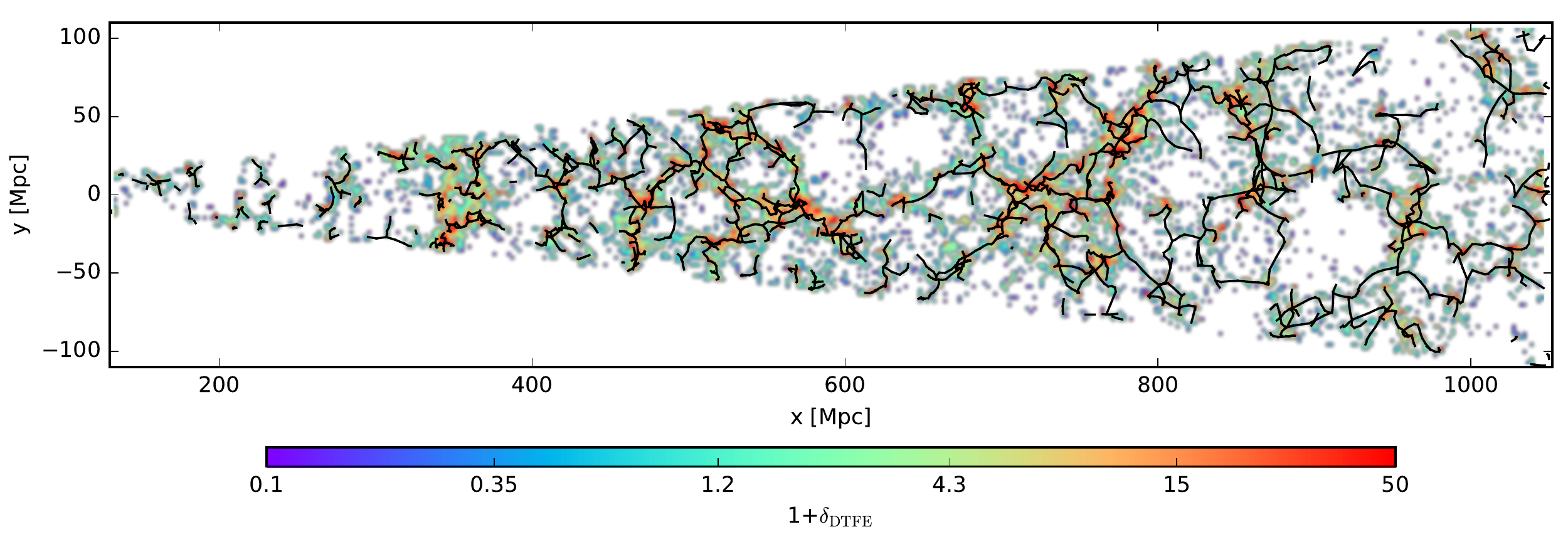}
\caption{Illustration of the filamentary network (black lines) extracted with the DisPerSE code within the $\pm$1.2$\degr$ of the central declination of the G12 field. 
The persistence threshold with which the filamentary network and the associated structures, used in this work and shown here, are extracted is 3$\sigma$.
Also shown is the density contrast of the underlying galaxy distribution, measured with the small-scale adaptive DTFE estimator (see text) and averaged over cells of 2.3 $\times$ 2.3 Mpc$^2$ (white colour is used for empty cells). In spite of the projection effects, the visual inspection reveals that filaments follow the ridges of the density field which connect the peaks together.
}
\label{Fig:skeleton}
\end{figure*}
This work uses the Discrete Persistent Structure Extractor \citep[DisPerSE; see][for illustrations in a cosmological context]{Sousbie2011b}, a geometric three-dimensional ridge extractor dealing directly with discrete datasets, making it particularly well adapted for astrophysical applications. It allows for a scale and parameter-free coherent identification of the 3D structures of the cosmic web  as dictated by the large-scale topology.
For a detailed description of the DisPerSE algorithm and its underlying theory, the reader is referred to \cite{Sousbie2011a};  its main features are summarised below.

DisPerSE is based on discrete Morse and persistence theories.
The Delaunay tessellation is used to generate a simplicial complex, i.e. a triangulated space with a geometric assembly of cells, faces, edges and vertices mapping the whole volume.  The Delaunay Tessellation Field Estimator \citep[DTFE;][]{Schaap2000, cautun2011} allows for estimating the density field at each vertex of the Delaunay  complex.  
The Morse theory enables to extract from the density field the critical points, i.e. points with a vanishing (discrete) gradient of the density field (e.g. maxima, minima and saddle points). These critical points are connected via the field lines tangent to the gradient field in every point. They induce a geometrical segmentation of space, where all the field lines have the same origin and destination, known as the Morse complex. This segmentation defines distinct regions called ascending and descending k-manifolds\footnote{the index k refers to the critical point the field lines emanate from (ascending) or converge to (descending), and is defined as its number of negative eigenvalues of the Hessian: a minimum of the field has index 0, a maximum has index 3 and the two types of saddles have index 1 and 2}. The morphological components of the cosmic web are then identified from these manifolds:  ascending 0-manifolds trace the voids, ascending 1-manifolds trace the walls and filaments correspond to the ascending 2-manifolds with its extremities plugged onto the maxima (peaks of the density field).
In addition to its ability to work with sparsely sampled data sets while assuming nothing about the geometry or homogeneity of the survey, 
DisPerSE allows for the selection of retained structures on the basis of the significance of the topological connection between critical points.
DisPerSE relies on persistent homology theory to pair  critical points according to the birth and death of a topological feature in the excursion set. 
The ``persistence" of a feature or its significance is assessed by the density contrast of the critical pair chosen to pass a certain signal-to-noise threshold. The noise level is defined relative to the RMS of persistence values obtained from random sets of points.
This thresholding eliminates less significant critical pairs, allowing to simplify the Morse complex, retaining its most topologically robust features. 
Figure~\ref{Fig:walls} shows that filaments outskirt walls, themselves circumventing voids. 
The filaments are made of a set of connected segments and their end points are connected to the maxima, the peaks of the density field where most of clusters and  large groups reside. Each wall is composed of the facets of tetrahedra from the Delaunay tessellation belonging to the same ascending 2-manifold.
In this work, DisPerSE is run on the flux-limited GAMA data with a 3$\sigma$ persistence threshold.
Figure~\ref{Fig:skeleton} illustrates the filaments for the G12 field, overplotted on the density contrast of the underlying galaxy distribution, $1 + \delta$, where the local density is estimated using the DTFE density estimator.
Even in this 2D projected visualisation one can see that filaments trace the ridges of the 3D density field connecting the density peaks between them.  
%

\subsection{Cosmic web metric}
\label{subsec:metric}

\begin{figure}
\includegraphics[width=8cm]{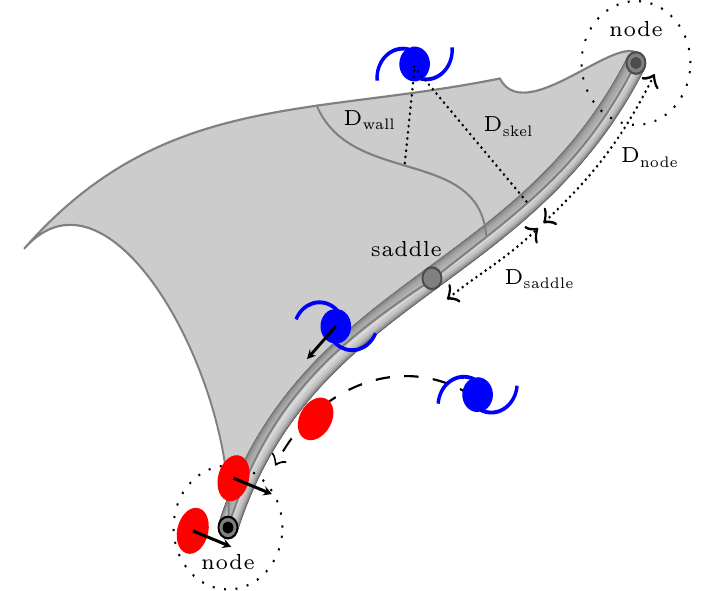}
\caption{Schematic view of the ``cosmic web'' metric in which the analysis is  performed. The position of a galaxy within the cosmic web is parametrised by its distance to the closest filament, $D_{\mathrm{skel}}$, and its distance to the closest wall, $D_{\mathrm{wall}}$. $D_{\mathrm{node}}$ and $D_{\mathrm{saddle}}$ represent the distances from the impact point to the node and saddle along the corresponding filament, respectively. 
}
\label{Fig:sketch_filament}
\end{figure}

Having identified the major cosmic web features, let us now define a new metric to characterise the environment of a galaxy, that will be referred to as the ``cosmic web metric'' and into which galaxies are projected.
Figure~\ref{Fig:sketch_filament} gives a schematic view of this framework. 
Each galaxy is assigned the distance to its closest filament, $D_{\mathrm{skel}}$.  
The impact point in the filament is then used to define the distances  along the filament toward the node, $D_{\mathrm{node}}$ and toward the saddle point,  $D_{\mathrm{saddle}}$.
Similarly, $D_{\mathrm{wall}}$ denotes the distance of the galaxy to its closest wall. 
In the present work, only distances $D_{\mathrm{node}}$, $D_{\mathrm{skel}}$ and $D_{\mathrm{wall}}$ are used. 
Other investigations of the environment in the vicinity of the saddle points are postponed to a forthcoming work.

The accuracy of the reconstruction of the cosmic web features is sensitive to the sampling of the dataset. The lower the sampling the larger the uncertainty on the location of the individual components of the cosmic web.
To account for the variation of the sampling throughout the survey, unless stated differently, all the distances are normalised by the redshift dependent mean inter-galaxy separation $\langle D_{\mathrm{z}} \rangle$, defined as $\langle D_{\mathrm{z}} \rangle \equiv n(z)^{-1/3}$, where $n(z)$ represents the number density of galaxies at a given redshift $z$. For the combined three fields of GAMA survey,  $\langle D_{\mathrm{z}} \rangle$ varies from 3.5 to 7.7 Mpc across the redshift range $0.03 \leq z \leq 0.25$, with a mean value of $\sim 5.6$ Mpc.

\section{Galaxy properties within the  cosmic web} 
\label{sec:mass_segregation}

In this section, the dependence of various galaxy properties, such as stellar mass, $u - r$ colour, \ssfr and type, with respect to their location within the cosmic web is analysed. 
First, the impact of the nodes, representing the largest density peaks, is investigated. 
Next, by excluding these regions, galaxy properties are studied within the intermediate density regions near the filaments.
Finally, the analysis is extended to the walls.

\subsection{The role of nodes via the red fractions}
\label{subsec:red_frac}

\begin{figure}
\includegraphics[width=\columnwidth]{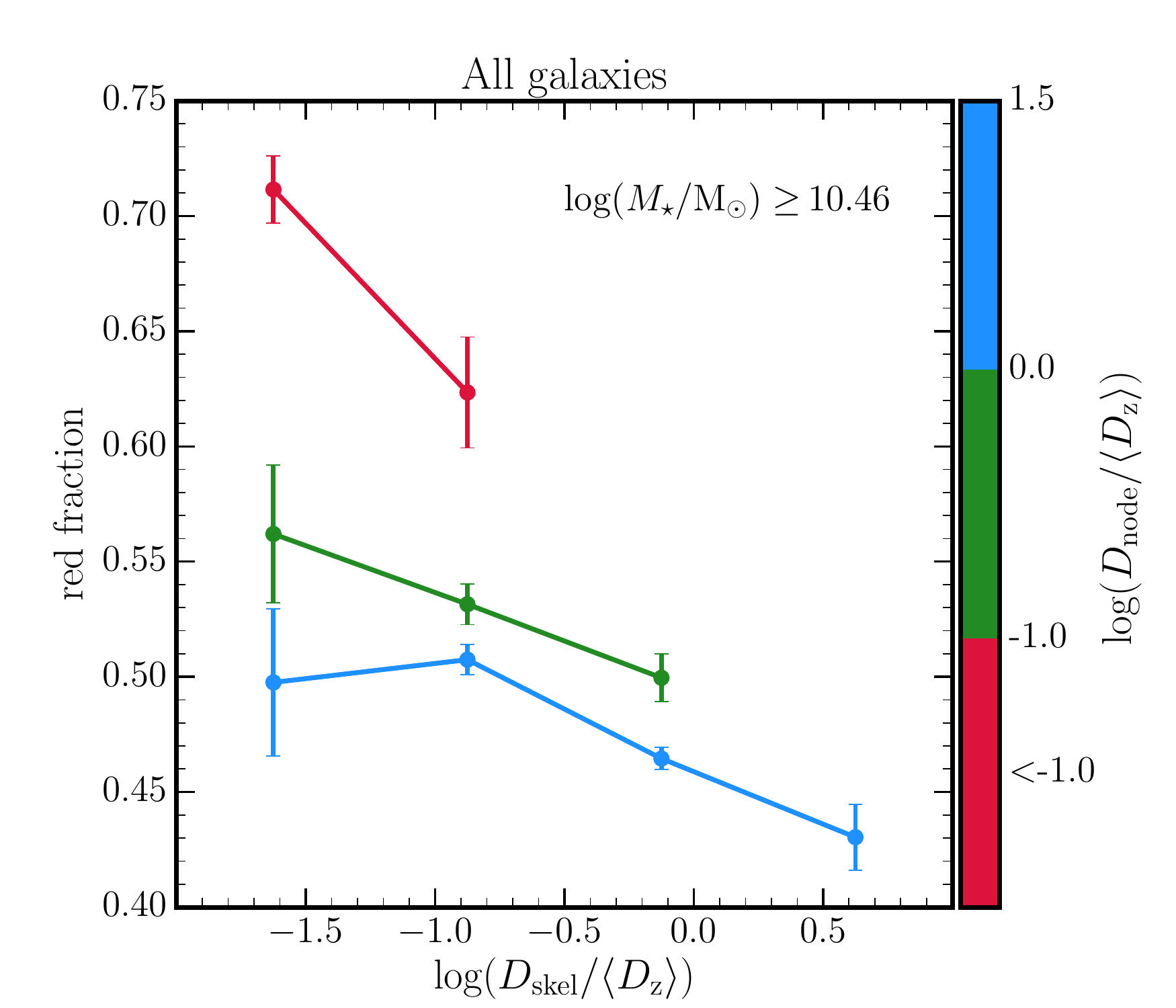}
\caption{Red fraction of galaxies (number of quiescent galaxies over the entire population) as a function of $D_{\mathrm{skel}}$ for three different bins of $D_{\mathrm{node}}$ as indicated by the colour. Both distances are normalised by the redshift-dependent mean inter-galaxy separation $\langle D_{\mathrm{z}} \rangle$.
Only galaxies with $\log(\Mstardot) \geq 10.46$ are considered. Star-forming and quiescent populations are matched in mass (see Section~\ref{subsec:grad_skel}). The error bars are calculated from 100 bootstrap samples. The fraction of red galaxies is found to increase with decreasing distances both to the closest filament $D_{\mathrm{skel}}$ and to the node of this $D_{\mathrm{node}}$. 
Recalling that $\langle D_{\mathrm{z}} \rangle \sim$ 5.6 Mpc, the fraction of passive galaxies increases at given $D_{\rm skel}$ by $\sim 20\%$ from several tens of Mpc away from the nodes (blue line) to less than $\sim$ 0.5 Mpc (red line). At fixed $D_{\rm node}$, the increase of the red fraction with decreasing distance to filaments is milder, of $\sim$10\% regardless of the distance to the node.
}
\label{Fig:red_frac}
\end{figure}

Let us start by analysing the combined impact of nodes and filaments on galaxies through the study of the red fractions.
The red fraction, defined as the number of passive galaxies with respect to the entire population, is analysed as a function of the distance to the nearest filament, $D_{\rm skel}$ and the distance to its associated node, $D_{\rm node}$.

This analysis is restricted to galaxies more massive than $\log(\Mstardot) \geq 10.46$, as imposed by the mass limit completeness of the passive population (see Section~\ref{sec:data}).  
The stellar mass distributions of the passive and star-forming populations are not identical, with the passive galaxies dominating the high mass end.  
Therefore, to prevent biases in the measured gradients introduced by such differences, the mass-matched samples are used. The detailed description of the mass-matching technique can be found in Appendix~\ref{subsec:appendix_mmatching}.

In Figure \ref{Fig:red_frac} the red fraction of galaxies is shown as a function of $D_{\rm skel}$ in three different bins of $D_{\rm node}$.  
While the fraction of passive galaxies is found to increase with decreasing distances both to the filaments and nodes, the dominant effect is  the distance to the nodes. 
At fixed $D_{\rm skel}$, the fraction of passive galaxies sharply increases with decreasing distance to the nodes. Recalling that the mean inter-galaxy separation $\langle D_{\mathrm{z}} \rangle \sim$ 5.6 Mpc, a 20 to 30\% increase in the fraction of passive galaxies is observed from several Mpc away from the nodes to less than $\sim$ 500 kpc. This behaviour is expected since the nodes represent the loci where most of the groups and clusters reside and reflects the well known colour-density \citep[e.g.][]{Blanton2003,Baldry2006,Bamford2009} and star formation-density \citep[e.g.][]{Lewis2002,Kauffmann2004} relations.
However, the gradual increase suggests that some physical processes already operate before the galaxies reach the virial radius of massive halos. 
At fixed $D_{\rm node}$, the fraction of passive galaxies increases with decreasing distance to filaments, but this increase is milder compared to that with respect to nodes:
an increase of $\sim$10\% is observed regardless of the distance to the nodes. 
These regions with intermediate densities appear to be a place where the transformation of galaxies takes place as emphasised in the next section.

\subsection{The role of filaments} 
\label{subsec:fil}

In order to infer the role played by filaments alone in the transformation of galactic properties, the impact of nodes, the high density regions has to be mitigated. By construction, nodes are at the intersection of filaments: they drive the well known galaxy type-density as well as stellar mass-density relations. To account for this bias, \citet[][]{Gay2010} and \citet[][]{Malavasi2017} adopted a method where a given physical property or distance of each galaxy was down-weighted by its local density.     
\cite{Laigle2017} adopted a more stringent approach by rejecting all galaxies that are too close to the nodes. This method allows to minimise the impact of nodes, avoiding the difficult-to-quantify uncertainty of the residual contribution of the density weighting scheme. 
The latter approach is therefore adopted.
As shown in Appendix~\ref{subsec:appendix_filaments}, this is achieved by rejecting all galaxies below a distance of 3.5 Mpc from a node.

\subsubsection{Stellar mass gradients}
\label{subsec:grad_skel}

\begin{figure*}
\includegraphics[width=0.85\textwidth]{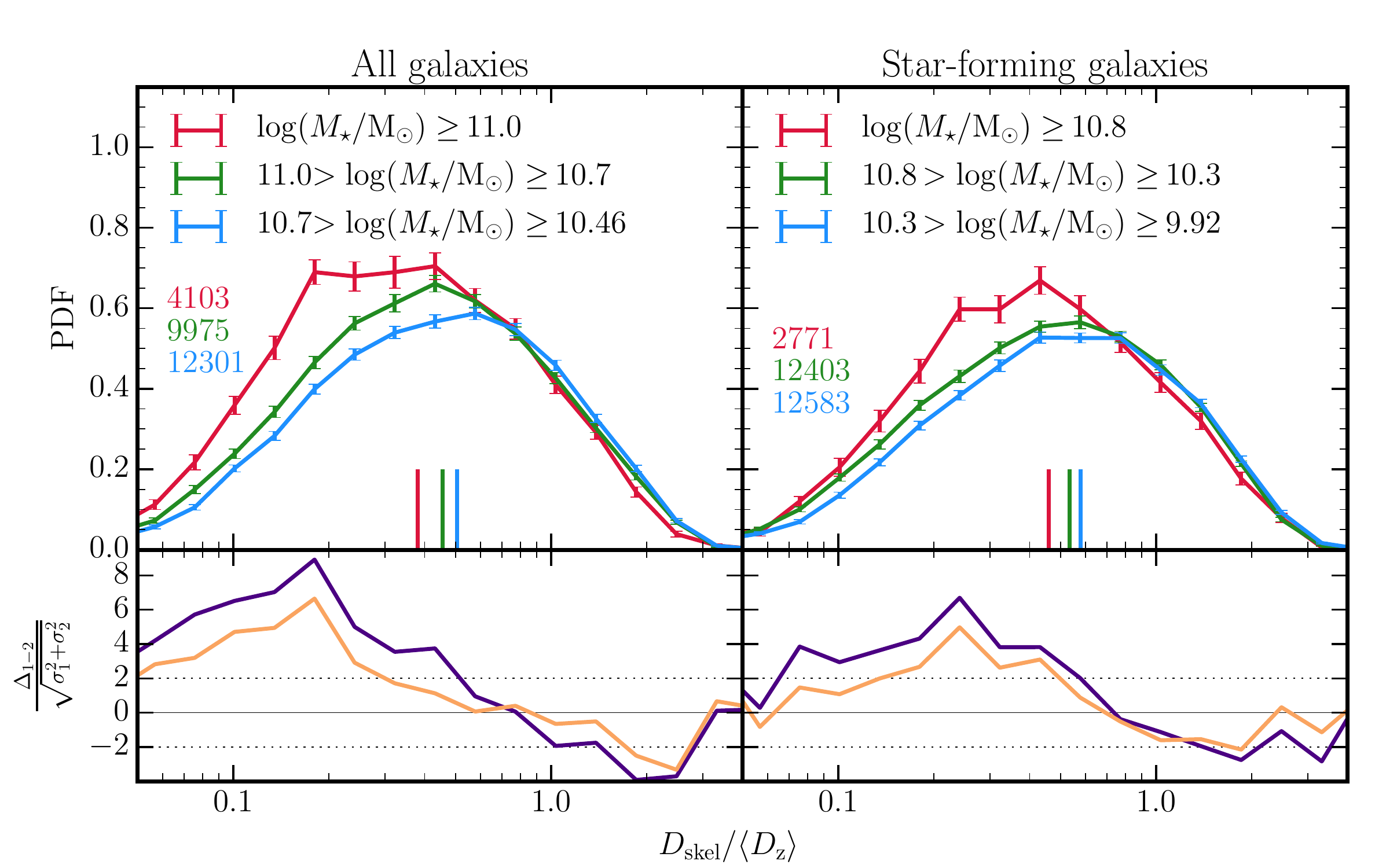} 
\caption{\textit{Top row:} Differential distributions of the distances to the nearest filament, $D_{\mathrm{skel}}$ (normalised by $\langle D_{\mathrm{z}} \rangle$, the redshift-dependent mean inter-galaxy separation) for the entire galaxy population (left panel) and star-forming galaxies alone (right panel) in three different stellar mass bins. 
Note that these bins are different for the two populations: this is due to the stellar mass completeness limit that is different (see Section~\ref{sec:data}). 
To highlight an effect specific to the filaments, the contribution of nodes is minimised (see text for details).
The vertical lines indicate the medians of the distributions and their values together with associated error bars are listed in Table~\ref{tab:medians_dskel_dwall}. 
The numbers of galaxies in different considered bins are indicated in each panel. The error bars are calculated from 100 bootstrap samples. 
There is a mass segregation of galaxies with respect to filaments of the entire as well as star-forming population: more massive galaxies tend to be preferentially located closer to the filaments compared to their lower-mass counterparts.
\textit{Bottom row:} Residuals in units of $\sigma$ between the two most extreme mass bins (purple line; $10.7 > \log (\Mstardot) \geq 10.46$ and $\log(\Mstardot) \geq 11.0$ on the left and $10.3 > \log (\Mstardot) \geq 9.92$ and $\log(\Mstardot) \geq 10.8$ on the right panels), and between the high and intermediate mass bins (orange solid line; $\log(\Mstardot) \geq 11.0$ and $11.0 > \log (\Mstardot) \geq 10.7$ on the left and $\log(\Mstardot) \geq 10.8$ and $10.8 > \log (\Mstardot) \geq 10.3$ on the right panels).
}
\label{Fig:PDF_dskel_masses}
\end{figure*}

\begin{table*}
\begin{threeparttable}
\caption{Medians for the PDFs displayed in Figures~\ref{Fig:PDF_dskel_masses} -- ~\ref{Fig:PDF_dwall_type}.}
\label{tab:medians_dskel_dwall}
\begin{tabular*}{0.8\textwidth}{@{\extracolsep{\fill}}lcccc}
\hline
\hline
& \multirow{2}{*}{selection\tnotex{tnote:panels}} & \multirow{2}{*}{bin} &  \multicolumn{2}{c}{median\tnotex{tnote:median}} \\
& & & $D_{\mathrm{skel}}/\left<D_\mathrm{z}\right>$  & $D_{\mathrm{wall}}/\left<D_\mathrm{z}\right>$\\
\hline
\hline
\multirow{6}{*}{mass\tnotex{tnote:mass_grad}}& \multirow{3}{*}{all galaxies }& $\log \, (\Mstardot)  \geq 11 $ &  0.379 $\pm$ 0.009 &  0.334 $\pm$ 0.005 \\
                  &                & $ 11 > \log \, (\Mstardot) \geq 10.7 $& 0.456 $\pm$ 0.007 &  0.381 $\pm$ 0.004 \\
                   &               & $ 10.7 > \log \, (\Mstardot) \geq 10.46$ & 0.505 $\pm$ 0.006 &  0.403 $\pm$ 0.004 \\
\cline{2-5}
& \multirow{3}{*}{SF galaxies}&$ \log \, (\Mstardot)  \geq 11$ & 0.459 $\pm$ 0.012 &  0.385 $\pm$ 0.011 \\
 &                                 &$11 > \log \, (\Mstardot) \geq 10.4$ & 0.534 $\pm$ 0.007 &  0.429 $\pm$ 0.006 \\
  &                                &$10.4 > \log \, (\Mstardot) \geq 9.92$ & 0.578 $\pm$ 0.007 & 0.453 $\pm$ 0.007 \\
\hline
\multirow{2}{*}{type\tnotex{tnote:type_grad}} &\multirow{2}{*}{SF vs passive\tnotex{tnote:SF_passive}}& star-forming& 0.504 $\pm$ 0.008 & 0.411 $\pm$ 0.006 \\
 &                                 & passive& 0.462 $\pm$ 0.007 & 0.376 $\pm$ 0.006 \\                                  
\hline
\end{tabular*}
\begin{tablenotes}
     \item\label{tnote:panels} panels of Figures ~\ref{Fig:PDF_dskel_masses} -- ~\ref{Fig:PDF_dwall_type}
     \item\label{tnote:median} medians of distributions as indicated in Figures~\ref{Fig:PDF_dskel_masses} -- ~\ref{Fig:PDF_dwall_type} by a vertical lines; 
              errors represent half width at half maximum of the bootstrap distribution, i.e. the distribution of medians from each of 100 bootstrap samples, fitted by a Gaussian curve
     \item\label{tnote:mass_grad} Figures~\ref{Fig:PDF_dskel_masses} and~\ref{Fig:PDF_dwall_masses}                                       
     \item\label{tnote:type_grad} Figures~\ref{Fig:PDF_dskel_type} and~\ref{Fig:PDF_dwall_type}     
     \item\label{tnote:SF_passive} only galaxies with stellar masses  $\log \, (\Mstardot) \geq 10.46$ are considered
    \end{tablenotes}
\end{threeparttable}
\end{table*}

Figure \ref{Fig:PDF_dskel_masses} shows the normalised probability distribution functions (PDFs) of the distance to the nearest filament $D_{\rm skel}$ in three stellar mass bins for the entire population and star-forming galaxies alone (top left and right panels, respectively). 
The medians of the PDFs, shown by vertical lines, are listed together with the corresponding error bars in Table~\ref{tab:medians_dskel_dwall}.
The significance of the observed trends is assessed by computing the residuals between the distributions in units of $\sigma$ (bottom panels), defined as $\Delta_{1-2}/\sqrt{\sigma^2_1+\sigma^2_2}$, where $\Delta_{1-2}$ is the difference between the PDFs of the populations $1$ and $2$, and $\sigma_1$ and  $\sigma_2$ are the corresponding standard deviations. 

For the entire population (left panels), differences between the PDFs of the three stellar mass bins are observed: the most massive galaxies ($\log(\Mstardot) \geq 11$) are located closer to the filaments than the intermediate population ($11 > \log(\Mstardot) \geq 10.7$), while the population with the lowest stellar masses ($10.7> \log(\Mstardot) \geq 10.46$) is found furthest away from the filaments. The significances of the difference between the most massive and the two lowest stellar mass bins are shown in the bottom panel.  Between the most extreme stellar mass bins (purple line), the difference exceeds 4$\sigma$ close to the filament and 2$\sigma$ at larger distances. It is slightly less significant 
between the intermediate and lowest stellar mass bins (orange line), but still in excess of 2$\sigma$ close to the filament. 
The differences between the PDFs can be also quantified in terms of their medians, where the differences between the highest and lowest stellar mass bins is significative at a $\sim 10 \sigma$ level (see Table~\ref{tab:medians_dskel_dwall}).
These results confirm previous claims of a mass segregation with respect to filaments, where the most massive galaxies are located near the core of the filaments, while the less massive ones tend to reside preferentially on their outskirts \citep[][]{Malavasi2017,Laigle2017}. 
As the impact of the nodes has been minimised, it is established that this stellar mass gradient is driven by the filaments themselves and not by the densest regions of the cosmic web. 

The mass segregation is also found among the star-forming population alone (right panels),  
such that more massive star-forming galaxies tend to be closer to the geometric core of the filament than their less massive counterparts. 
Note that the mass bins for star-forming galaxies differ from mass bins  used for the entire population. The completeness stellar mass limit allows us to decrease the lowest mass bin to $\log(\Mstardot) = 9.92$ when considering the star-forming galaxies alone (see Section~\ref{sec:data}).
The significance of these stellar mass gradients between the extreme stellar mass bins exceeds $4\sigma$ near the filaments, while the difference of the medians reaches a $\sim 8 \sigma$ level (see Table~\ref{tab:medians_dskel_dwall}).

\subsubsection{Type gradients}
\label{subsec:grad_type}

\begin{figure}
\includegraphics[width=\columnwidth]{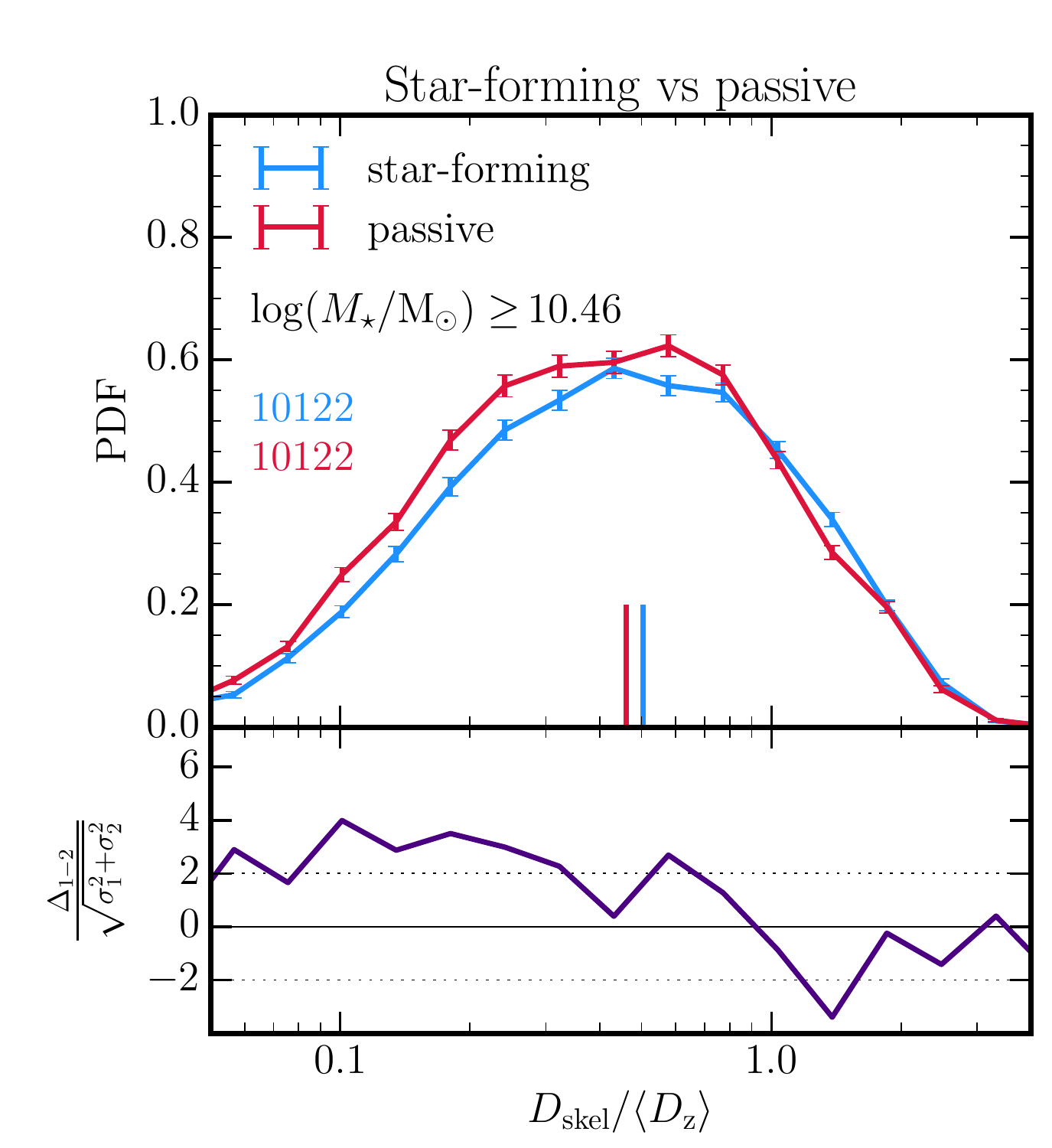} 
\caption{\textit{Top:} Differential distributions of the distances to the nearest filament, $D_{\mathrm{skel}}$ (normalised by $\langle D_{\mathrm{z}} \rangle$, the redshift-dependent mean inter-galaxy separation) for star-forming and quiescent galaxies that have been matched in mass (see text for details). 
To highlight an effect specific to the filaments, the contribution of node is minimised (see text for details).
The vertical lines indicate the medians of the distributions and their values together with associated error bars are listed in Table~\ref{tab:medians_dskel_dwall}. The numbers of galaxies in different considered bins are indicated in each panel. The error bars are calculated from 100 bootstrap samples.
Galaxies are found to segregate, relative to filaments, according to their type: quiescent galaxies tend to be preferentially located closer to the filaments compared to their star-forming counterparts. 
\textit{Bottom:} Residuals in units of $\sigma$  between the star-forming and passive galaxies.  
}
\label{Fig:PDF_dskel_type}
\end{figure}

Let us now investigate the impact of the filamentary network on the type/colour of galaxies. 
To do so, galaxies are split by type between star-forming and passive galaxies based on the dust corrected $u-r$ colour as discussed in Section~\ref{subsec:galaxies}. 
As for the analysis of the red fraction (Section~\ref{subsec:red_frac}), the sample is restricted to galaxies with $\log(\Mstardot) \geq 10.46$ and the star-forming and passive populations are matched in stellar mass.
Figure~\ref{Fig:PDF_dskel_type} shows the PDFs of the normalised distances $D_{\mathrm{skel}}$ within the mass-matched samples of star-forming and passive populations, which by construction have the same number of galaxies. 
Galaxies are found to segregate according to their type such that passive galaxies tend to reside in regions located closer to the core of filaments than their star-forming counterparts. 
The significance of the stellar mass gradients between the two populations exceeds $3\sigma$ near filaments while the difference between the medians reaches a $\sim 4\sigma$ level (see Table~\ref{tab:medians_dskel_dwall}).

\subsubsection{Star formation activity gradients}

To explore whether the impact of filaments on the star formation activity of galaxies can be detected beyond the red fractions and type segregation reported above, the focus is now on the star-forming population alone through the study of their (dust corrected) $u - r$ colour and \ssfr. 
 
\begin{figure}
\includegraphics[width=\columnwidth]{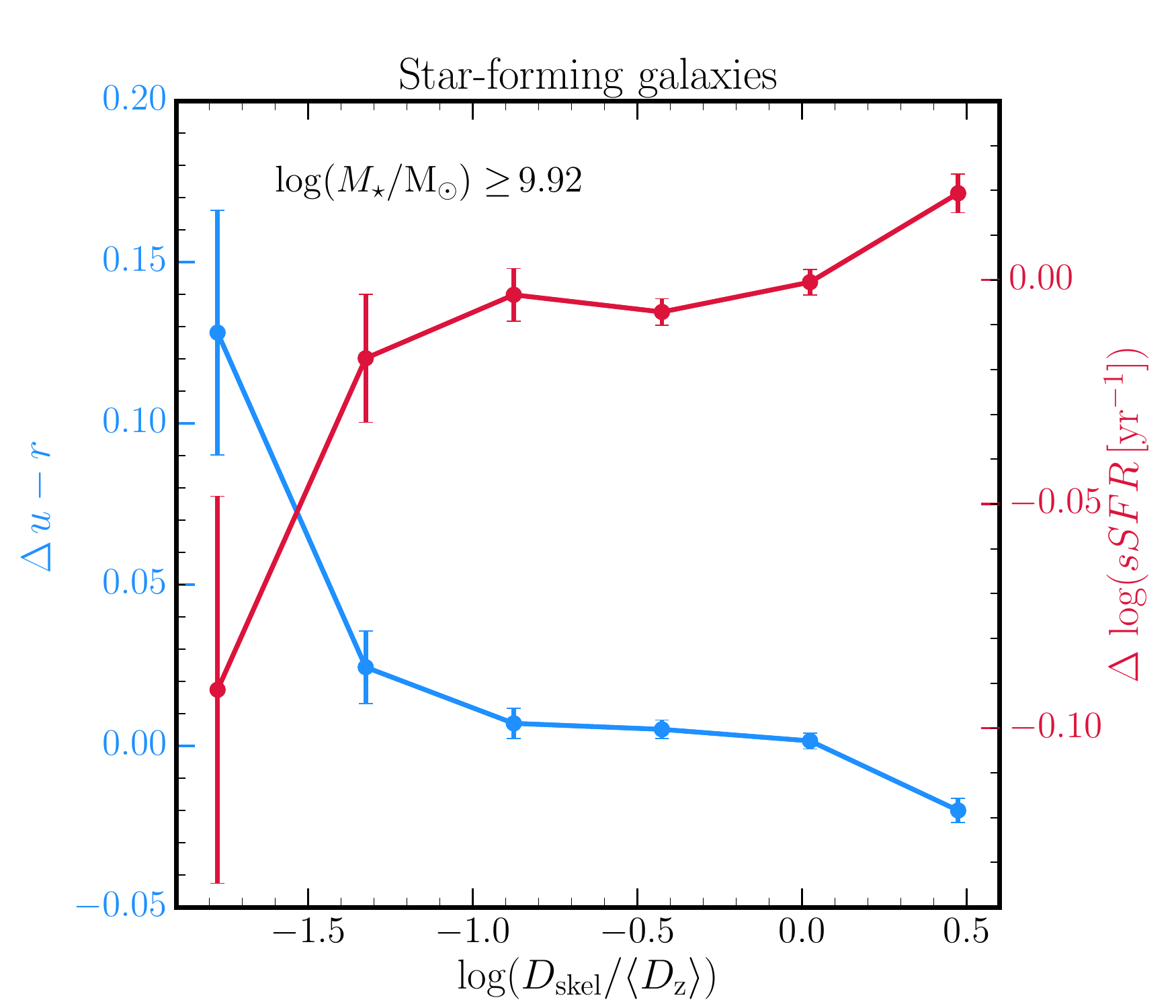}
\caption{$u - r$ colour (blue line) and \ssfr (red line) of star-forming galaxies as a function of of $D_{\mathrm{skel}}$. The y-axes indicate the amount by which $u-r$ colour and \ssfr differ from the median values at given mass (see text for details).  
Only galaxies with $\log(\Mstardot) \geq 9.92$ and faraway from nodes (at $D_{\rm node}  >  3.5$ Mpc) are considered. Star-forming galaxies tend to have higher $u -r$ colour (tend to be redder) and lower \ssfr when they get closer to the filaments than their more distant counterparts.}
\label{Fig:ssfr_u_r}
\end{figure}

Both these quantities are known to evolve with stellar mass which itself varies within the cosmic web (see above). 
To remove this mass dependence, the offsets of $u - r$ colour and \ssfr, $\Delta u - r$ and $\Delta \ssfr$ respectively, from the median values of all star-forming galaxies at a given mass are computed for each galaxy.  
Figure~\ref{Fig:ssfr_u_r} shows the medians of $\Delta u - r$ and $\Delta \ssfr$ as a function of $D_{\rm skel}$.  Both quantities are found to carry the imprint of the large-scale environment.  
At large distances from the filaments ($D_{\rm skel}\ge $ 5 Mpc), star-forming galaxies are found to be more active than the average. At intermediate distances (0.5$\le D_{\rm skel}\le$ 5 Mpc), star formation activity of star-forming galaxies do not seem to evolve with the distance to the filaments, 
while in the close vicinity of the filaments  ($ D_{\rm skel}\le$ 0.5 Mpc), they show signs of a decrease in star formation efficiency (redder colour and lower \ssfr). The significance of these results will be discussed in Section~\ref{sec:discussion}.
  
\subsection{The role of walls in mass and type gradients}
\label{subsec:grad_wall}

\begin{figure*}
\includegraphics[width=0.85\textwidth]{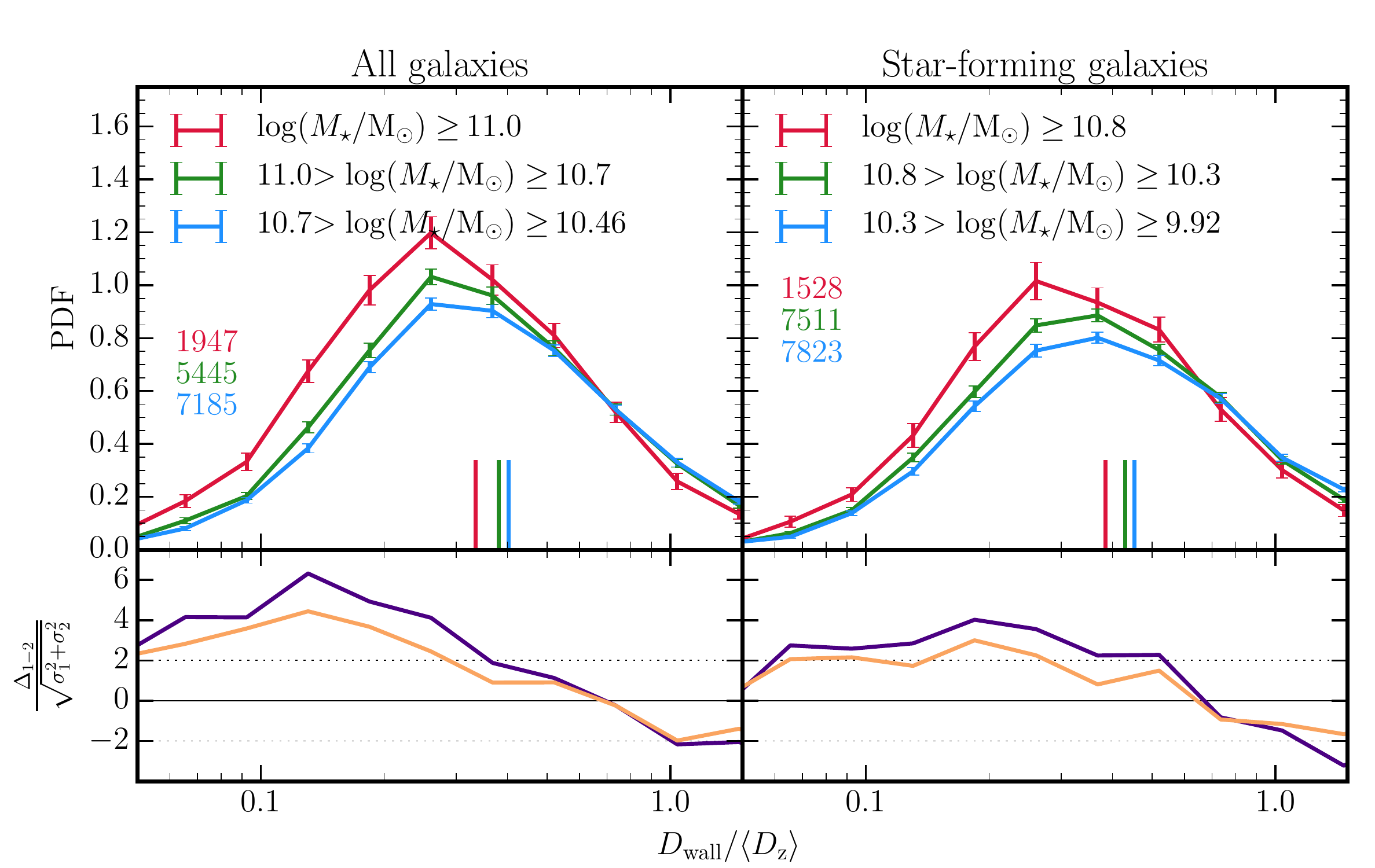}
\caption{\textit{Top row:} As in Figure~\ref{Fig:PDF_dskel_masses}, but for the distances to the nearest wall, $D_{\mathrm{wall}}$.
To minimise the contribution of nodes and filaments to the measured signal, galaxies located closer to a node than 3.5 Mpc and closer to a filament than 2.5 Mpc are removed form the analysis. 
There is a mass segregation of galaxies with respect to walls of the entire as well as star-forming population: more massive galaxies tend to be preferentially located closer to the filaments compared to their lower-mass counterparts.
\textit{Bottom row:} Residuals in units of $\sigma$ as in Figure~\ref{Fig:PDF_dskel_masses}. 
}
\label{Fig:PDF_dwall_masses}
\end{figure*}

\begin{figure}
\includegraphics[width=\columnwidth]{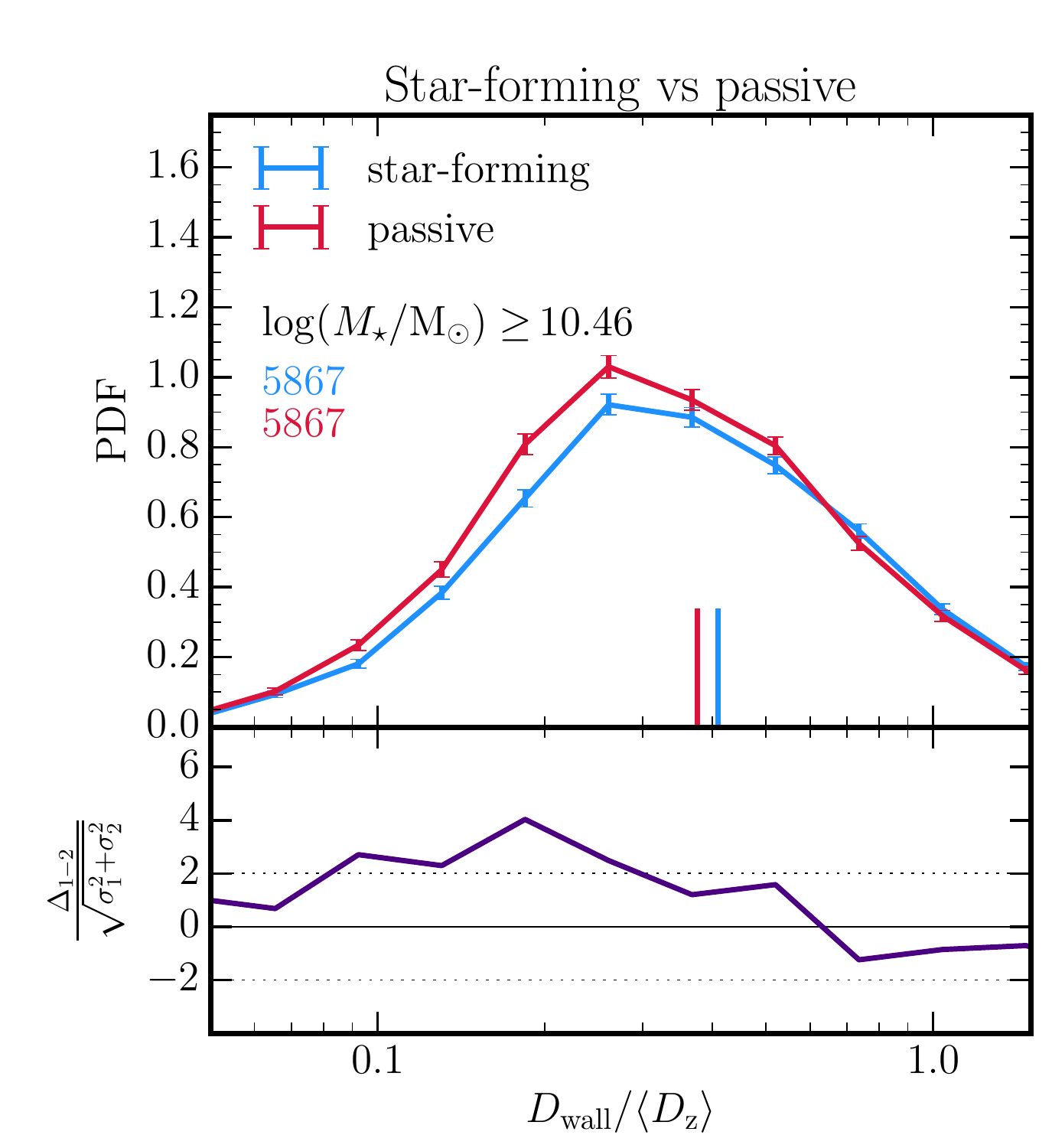}
\caption{\textit{Top row:}  As in Figure~\ref{Fig:PDF_dskel_type}, but for the distances to the nearest wall, $D_{\mathrm{wall}}$.
To minimise the contribution of nodes and filaments to the measured signal, galaxies located closer to a node than 3.5 Mpc and closer to a filament than 2.5 Mpc are removed form the analysis. 
Galaxies are found to segregate, with respect to walls, according to their type: quiescent galaxies tend to be preferentially located closer to the walls compared to their star-forming counterparts.
\textit{Bottom row:} Residuals in units of $\sigma$ as in Figure~\ref{Fig:PDF_dskel_type}. 
}
\label{Fig:PDF_dwall_type}
\end{figure}

Let us now investigate the impact of walls on galaxy properties. Figures~\ref{Fig:PDF_dwall_masses} and \ref{Fig:PDF_dwall_type} show the PDFs of the distances to the closest wall $D_{\rm wall}$ for the same selections as in Figures \ref{Fig:PDF_dskel_masses} and \ref{Fig:PDF_dskel_type}, respectively.
The distances are again normalised by the redshift dependent mean inter-galaxy separation $\langle D_{\mathrm{z}} \rangle$. The values of medians with corresponding error bars are listed in Table~\ref{tab:medians_dskel_dwall}. 
As for filaments, one seeks signatures induced by a particular environment solely, walls in this case.  
Given that filaments are located at the intersections between walls, in addition to the contamination by nodes, which is of concern for filaments, one has to make sure that the contribution of filaments themselves 
is minimised as well. Following the method adopted in Section~\ref{subsec:grad_skel}, Appendix~\ref{subsec:appendix_walls} shows that this can be achieved by removing from the analysis galaxies having distances to the nodes smaller than 3.5 Mpc and distances to the closest filaments less than 2.5 Mpc.   

The derived trends are qualitatively similar to those measured with respect to  filaments. Massive galaxies are located closer to walls compared to their low-mass counterparts; star-forming galaxies preferentially reside in the outer regions of walls; and mass segregation is present also among star-forming population of galaxies with more massive star-forming galaxies having smaller distances to the walls than their low-mass counterparts.
Since the walls typically embed smaller-scale filaments, the net effect of transverse gradients perpendicular to these filaments should add up to transverse 
gradients perpendicular to walls.

The significance of the measured trends, in terms of the residuals between medians (see Table~\ref{tab:medians_dskel_dwall}), is above 3$\sigma$ for all considered gradients, slightly lower than for the gradients towards filaments. The deviations of $\sim 10 \sigma$ and $\sim 5 \sigma$ are detected between the highest and lowest stellar mass bins among the whole and star-forming population alone, respectively, while between the star-forming and passive galaxies it reaches $\sim 4 \sigma$, as in the case of gradients towards filaments.

\section{Comparison with the Horizon-AGN simulation}
\label{sec:hagn}
 
\begin{figure*}
    \centering
    \begin{subfigure}{\textwidth}
        \centering
        \includegraphics[width=0.75\textwidth]{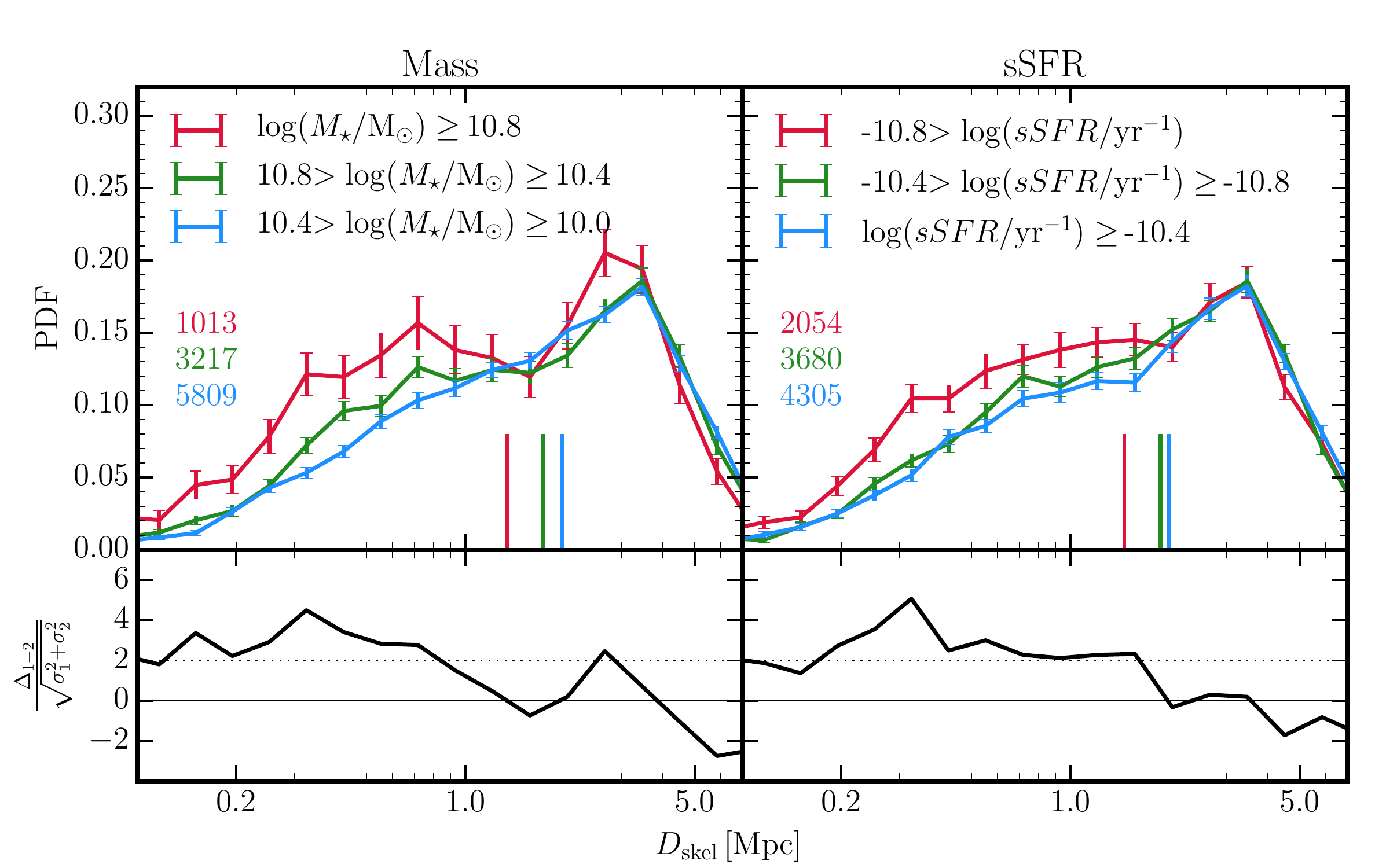}
        \caption{Differential distributions of the distances to the nearest filament, $D_{\mathrm{skel}}$.}
    \end{subfigure}%
    \\
    \begin{subfigure}{\textwidth}
        \centering
        \includegraphics[width=0.75\textwidth]{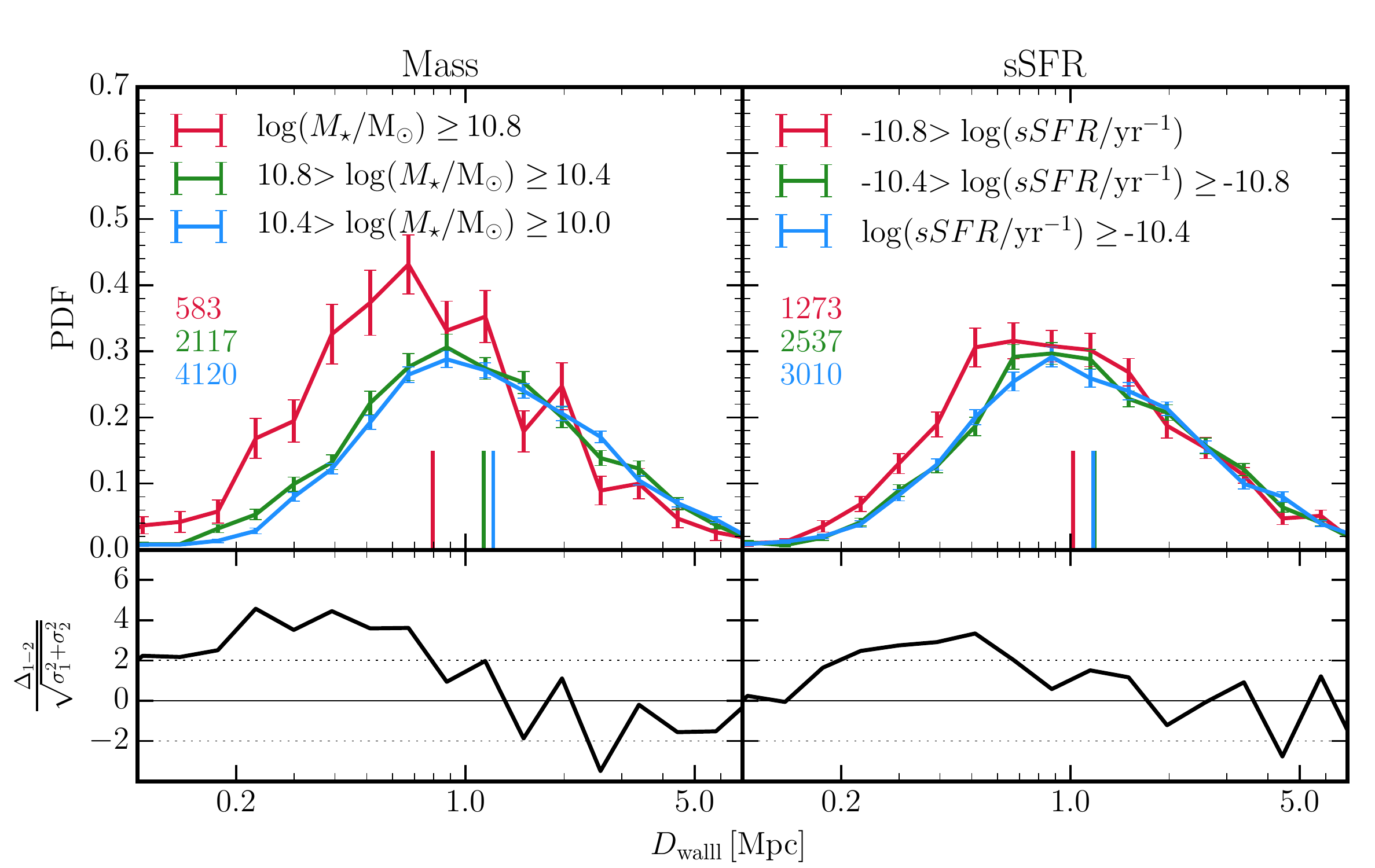}
        \caption{Differential distributions of the distances to the nearest wall, $D_{\mathrm{wall}}$.}
    \end{subfigure}
    \caption{\textit{Top rows:} Differential distributions of the distances as a function of stellar mass (left panels) and \ssfr (right panels) for galaxies in \hagn. 
To minimise the contribution of nodes and filaments to the measured signal, galaxies located closer to a node than 3.5 Mpc and closer to a filament than 1 Mpc are removed form the analysis. The vertical lines indicate the medians of the distributions (see Table~\ref{tab:medians_dskel_dwall_HzAGN} for the numerical values). Numbers of galaxies in different considered bins are indicated in each panel. 
There is mass and \ssfr segregation of galaxies with respect to both filaments and walls: more massive and less star-forming galaxies tend to be preferentially located closer to the cores of filaments and walls compared to their lower-mass and more star-forming counterparts, respectively. These results are in qualitative agreement with the measurements in GAMA.
 \textit{Bottom rows:} Residuals in units of $\sigma$ between the two most extreme mass and \ssfr bins, $\log(\Mstardot) \geq 10.8$ and $10.4 > \log (\Mstardot) \geq 10$ on the left 
and $-10.8 > \log (\ssfr/\mathrm{yr}^{-1})$ and $\log(\ssfr/\mathrm{yr}^{-1}) \geq -10.4$ on the right panels, respectively.
 }
    \label{Fig:PDF_dskel_dwall_HzAGN}
\end{figure*}

In this section, a qualitative support for the results on the mass and star-formation activity segregation is provided via the analysis of the large-scale cosmological hydrodynamical simulation \hagn~\citep{Dubois2014}. 
Note that the main purpose of such an analysis is to provide a reference measurement of gradients in the context of a large-scale "full physics" experiment.      
The construction of the GAMA-like mock catalogue is not performed because the geometry of \hagn\, does not allow us to recover the entire GAMA volume 
and the flux-limited sample requires a precise modelling of fluxes in different bands.   
 
A brief summary of some of the main features of the simulation can be found in Appendix~\ref{sec:appendix_hagn}.
Here, the results on the mass and \ssfr gradients towards filaments and walls are presented. 
The \hagn\,simulation is analysed at low redshift ($z \sim 0.1$), comparable to the mean redshift studied in this paper, and the same analysis is performed as in the GAMA data. The filamentary network and associated structures are extracted by running the DisPerSE code with the persistence threshold of $3 \sigma$.

Figure~\ref{Fig:PDF_dskel_dwall_HzAGN} shows the mass (left panels) and \ssfr (right panels) gradients towards filaments (figure a) and walls (figure b) as measured in the \hagn\,simulation. The impact of the nodes and filaments on the measured signal is minimised by removing from the analysis galaxies that are closer to the node than 3.5 Mpc and closer to the filament than 1 Mpc. The detailed description of the method used to identify these cuts in distances can be found in Appendix~\ref{subsec:appendix_filaments}. 
Consistently with the measurements in GAMA, galaxies in \hagn\, are found to segregate by  stellar mass, with more massive galaxies being preferentially  closer to both the filaments and walls than their low-mass counterparts. 
Similarly, the presence of the \ssfr gradient, whereby less star-forming galaxies tend to be closer to the cores of filaments and walls than their more star-forming counterparts, is in qualitative agreement with the type/colour gradients detected in the GAMA survey.   
Note that the three bins of \ssfr are used to separate out the highly star-forming galaxies, with $\log(\ssfr/\mathrm{yr}^{-1}) \geq -10.4$, from passive ones, with $\log (\ssfr/\mathrm{yr}^{-1}) < -10.8$, in order to compare with the type gradients in the observations. In the simulation, \ssfr is a more reliable parameter for type than the colour.

The significance of the trends is measured, as previously, in terms of the residuals between medians (see Table~\ref{tab:medians_dskel_dwall}). 
For the gradients towards filaments, the difference of $\gtrsim 6 \sigma$ is found between the most extreme, both mass and \ssfr, bins, while it drops to $\sim 2-3 \sigma$ between the intermediate and lowest bins.
For the gradients towards walls, the deviation between the most extreme bins is $\sim$ 10 and $4\sigma$ for mass and \ssfr bins, respectively, while there is only a little to no difference between intermediate and lowest stellar mass and \ssfr bins, respectively. The gradients are slightly less significant than in the GAMA measurements, most likely due to the low numbers of galaxies per individual bins in \hagn, but qualitatively similar as in GAMA.

\begin{table*}
\begin{threeparttable}
\caption{Medians for the PDFs displayed in Figure~\ref{Fig:PDF_dskel_dwall_HzAGN}.}
\label{tab:medians_dskel_dwall_HzAGN}
\begin{tabular*}{0.8\textwidth}{@{\extracolsep{\fill}}lccc}
\hline
\hline
 
\multirow{2}{*}{selection\tnotex{tnote:panels}} & \multirow{2}{*}{bin} &  \multicolumn{2}{c}{median\tnotex{tnote:median}} \\
& & $D_{\mathrm{skel}}$ [Mpc]  & $D_{\mathrm{wall}}$ [Mpc]\\
\hline
\hline
\multirow{3}{*}{Mass }& $\log \, (\Mstardot)  \geq 10.8 $ & 1.34 $\pm$ 0.09 &  0.79 $\pm$ 0.04 \\
                                 & $ 10.8 > \log \, (\Mstardot) \geq 10.4 $ & 1.73 $\pm$ 0.08 & 1.14 $\pm$ 0.03  \\
                                  & $ 10.4 > \log \, (\Mstardot) \geq 10 $ &1.97 $\pm$ 0.04  & 1.22 $\pm$ 0.02  \\
\hline
\multirow{3}{*}{\ssfr\tnotex{tnote:ssfr_grad}}&$ -10.8 > \log \, (sSFR/\mathrm{yr}^{-1})  $ & 1.46 $\pm$ 0.07 & 1.02 $\pm$ 0.03 \\
                                  &$ -10.4 > \log \, (sSFR/\mathrm{yr}^{-1}) \geq -10.8$ & 1.88 $\pm$ 0.06 & 1.18 $\pm$ 0.03  \\
                                  &$ \log \, (sSFR/\mathrm{yr}^{-1}) \geq -10.4$ & 2.0 $\pm$ 0.04 & 1.18 $\pm$ 0.02  \\                              
\hline
\end{tabular*}
\begin{tablenotes}
     \item\label{tnote:panels} panels of Figure~\ref{Fig:PDF_dskel_dwall_HzAGN}
     \item\label{tnote:median} medians of distributions as indicated in Figure~\ref{Fig:PDF_dskel_dwall_HzAGN} by a vertical lines; errors are computed as in Table~\ref{tab:medians_dskel_dwall} 
     \item\label{tnote:ssfr_grad} only galaxies with stellar masses  $\log \, (\Mstardot) \geq 10$ are considered
\end{tablenotes}
\end{threeparttable}
\end{table*}


\section{The relative impact of density}
\label{sec:density}
  
Let us now address the following questions:
what is the specific role of the geometry of the large-scale environment in establishing mass and type/colour large-scale gradients? Are these gradients driven solely by density, or does the large scale anisotropy of the cosmic web provide a specific signature?

A key ingredient in answering these questions is the choice of the scale at which the density is inferred.
The properties of galaxies at a given redshift are naturally a signature of their past lightcone. This lightcone in turn correlates  with the galaxy's environment: the larger the scale is, the longer the look-back time one must consider, the more integrated the net effect of this environment. This past environment accounts for the total accreted mass of the galaxy, but may also impact the geometry of the accretion history and more generally other galactic properties such as its star formation efficiency, its colour or its spin.
At small scales,  the density correlates with 
the most recent and stochastic processes, while going to larger scales allows taking the integrated hence smoother history of galaxies into account. 
Since this study is concerned about the statistical impact of the large scale structure on  galaxies, it is natural to consider scales large enough to average out local recent events they may have encountered, such as binary interactions, mergers, outflows.
Therefore in the discussion below, the density is computed at the scale of 8 Mpc, the "smallest" scale at which the effect of the anisotropic large-scale tides can be detected.
  
In practice, in order to try to disentangle the effect of density from that of the anisotropic large-scale tides, the following reshuffling method \citep[e.g.][]{Malavasi2017} is adopted. For mass gradients, ten equipopulated density bins are constructed and in each of them the stellar masses of galaxies are randomly permuted.
By construction, the underlying mass-density relation is preserved, but this procedure randomises the relation between the stellar mass and the distance to the filament or the wall. 
For the type/colour gradients, in each of ten equipopulated density bins, ten equipopulated stellar mass bins are constructed. Within each of such bins, $u - r$ colour of galaxies are randomly permuted. Thus by construction, this preserves the underlying colour-(mass)-density relation, but  breaks the relation between the colour/type and the distance to the particular environment, the filament or wall.    

In order to account for the variation of the density through the survey, the density contrast, defined as $1 + \delta = n/ n(z)$, where $n(z)$ corresponds to the mean redshift dependent number density, is used in logarithmic bins. The number density $n$ is computed in the Gaussian kernel and
every time five reshuffled samples are constructed.

\begin{figure*}
    \centering
    \begin{subfigure}{\textwidth}
        \centering
        \includegraphics[width=\textwidth]{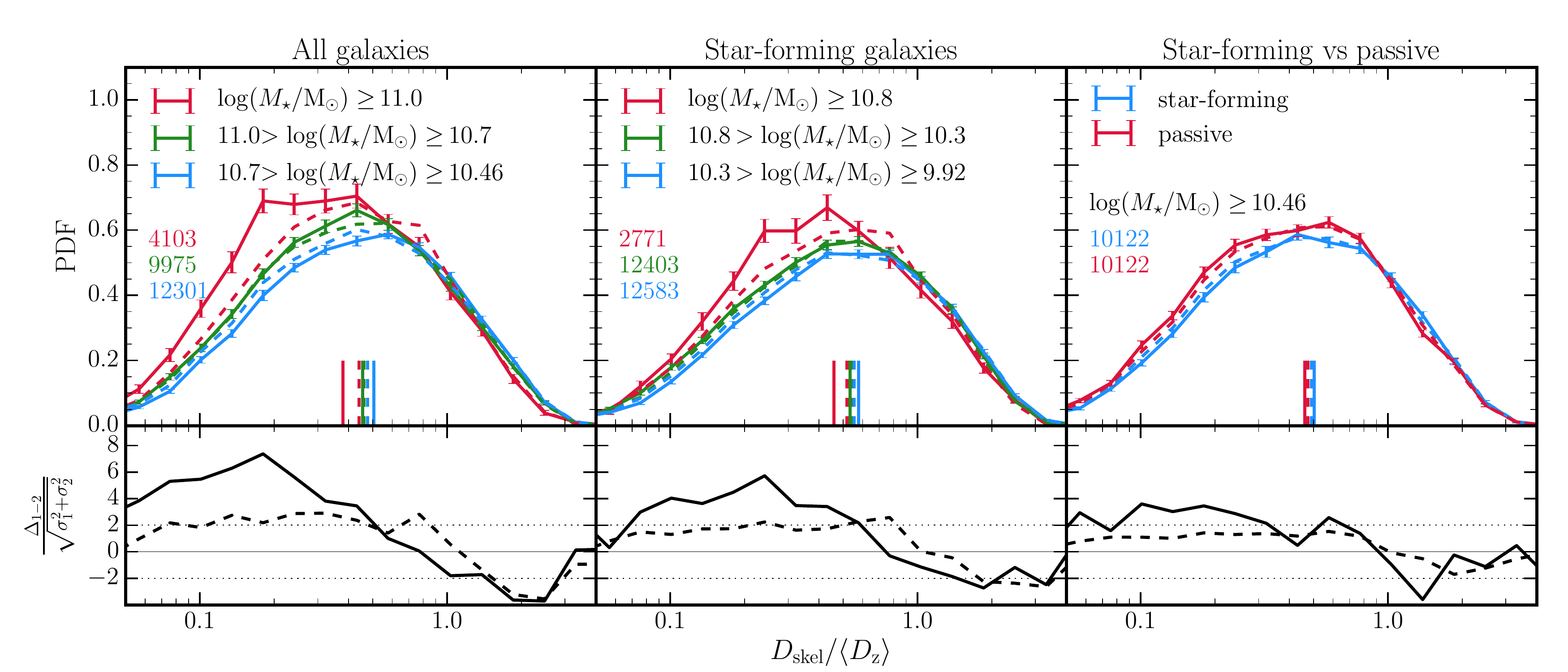}
        \caption{Reshuffling.}
    \end{subfigure}%
    \\
    \begin{subfigure}{\textwidth}
        \centering
        \includegraphics[width=\textwidth]{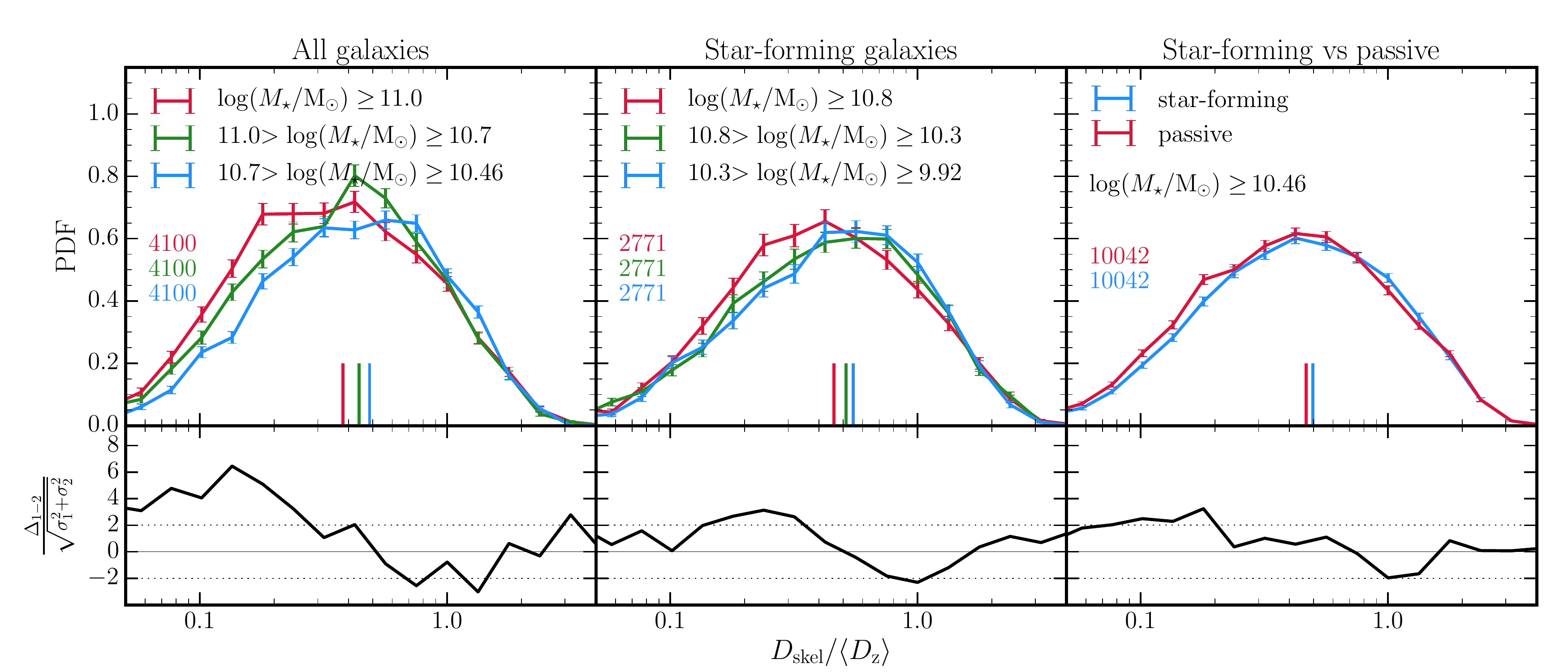}
        \caption{Density matching.}
    \end{subfigure}
    \caption{\textit{Top rows:} Differential distributions of the normalised distances to the nearest filament, $D_{\mathrm{skel}}$ as a function of stellar mass of the entire galaxy population (left panels), for star-forming galaxies only (middle panels) and as a function of galaxy's type (right panels) with reshuffling (Figure a) and with density-matched samples (Figure b). In Figure (a), the distributions before applying the reshuffling method (solid lines) are compared to the results after the reshuffling (dashed lines). 
Figure (b) illustrates the distributions for the galaxy samples that are matched so that their density distributions are the same (see the text for details on the matching). 
The density estimators used both in the reshuffling and density matching corresponds to the (large-scale) density computed in the Gaussian kernel. at the scale of 8 Mpc.  
As previously, the contribution of nodes to the measured signal is minimised.
The numerical values of medians, shown as a vertical lines, are listed in Table~\ref{tab:medians_dskel_G8Mpc_reshuff_dmatching}. 
The two methods yield qualitatively similar result: on the one hand when the large-scale density is used in reshuffling, the signal is reduced (dashed lines, Figure a) suggesting that the measured gradients (solid lines, Figure a) are not driven by the density at this scale, on the other hand, the gradients are measured within the samples that are matched in density at large-scale. 
 \textit{Bottom rows:} Residuals in units of $\sigma$ between the highest and lowest mass bins (left and middle panels) and between the star-forming and passive galaxies (right panels).
 }
    \label{Fig:PDF_dskel_G8Mpc_reshuff_dmatching}
\end{figure*}

In Figure~\ref{Fig:PDF_dskel_G8Mpc_reshuff_dmatching} (a), the mass and type gradients towards filaments, as measured in GAMA and previously shown in Figures~\ref{Fig:PDF_dskel_masses} and \ref{Fig:PDF_dskel_type}, are compared with the outcome of the reshuffling technique. The original signal is found to be substantially reduced after the reshuffling of masses and colours of galaxies. For the mass gradients, the deviation between the most extreme bins before reshuffling exceeds $3 \sigma$, while after the reshuffling, the signal gets reduced, with typical deviations of $\sim 1 \sigma$. The original signal for the type/colour gradients is weaker than in the case of the mass gradients, however, it is similarly nearly cancelled out once the reshuffling method is applied. 
The values of medians of the distributions after the reshuffling can be found in Table~\ref{tab:medians_dskel_G8Mpc_reshuff_dmatching}. Qualitatively similar behaviour is obtained for the gradients towards walls (not shown here).   
The analysis in \hagn\, provides a qualitative support for these results. In Appendix~\ref{subsec:appendix_density_HzAGN}, Figure~\ref{Fig:PDF_dskel_reshuff_HzAGN} (a), the same reshuffling method is applied to simulated galaxies. The density used for this test is computed in the Gaussian kernel at 5 Mpc. This scale corresponds to the $\sim 1.5 \times$ mean inter-galaxy separation in \hagn, consistently with the GAMA data.

Alternatively, to assess the impact of the density on the measured gradients within the cosmic web, one may want to use density matching. The purpose of this method is to construct mass- and colour-density matched samples, whereby galaxies with different masses and/or colours have similar density distributions, in order to make sure that the measured properties are not driven by their differences (see Appendix~\ref{subsec:appendix_dmatching} for details on the matching technique). 
As shown in Figure~\ref{Fig:PDF_dskel_G8Mpc_reshuff_dmatching} (b), the main result on the density-matching technique leads to the same conclusions as the reshuffling method. 
After matching galaxy populations in the large-scale density, mass and type gradients towards filaments and walls are still detected, suggesting that beyond the density, large scale structures of the cosmic web do impact  these galactic properties.

\begin{table*}
\begin{threeparttable}
\caption{Medians for the PDFs displayed in Figure~\ref{Fig:PDF_dskel_G8Mpc_reshuff_dmatching}: large-scale density}
\label{tab:medians_dskel_G8Mpc_reshuff_dmatching}
\begin{tabular*}{0.8\textwidth}{@{\extracolsep{\fill}}lccccc}
\hline
\hline
& \multirow{3}{*}{selection\tnotex{tnote:panels}} & \multirow{3}{*}{bin} &  \multicolumn{3}{c}{median\tnotex{tnote:median}} \\
& & &  \multicolumn{3}{c}{$D_{\mathrm{skel}}/\left<D_\mathrm{z}\right>$ } \\
\cline{4-6}
& & & original\tnotex{tnote:before}  & reshuffling\tnotex{tnote:after} & matching \tnotex{tnote:dmatching} \\
\hline
\hline
\multirow{6}{*}{Masses}& \multirow{3}{*}{All galaxies }& $\log \, (\Mstardot)  \geq 11 $ &  0.379 $\pm$ 0.009 &  0.441 $\pm$ 0.009 & 0.379 $\pm$ 0.01\\
                  &                & $ 11 > \log \, (\Mstardot) \geq 10.7 $& 0.456 $\pm$ 0.007 &  0.463 $\pm$ 0.006 & 0.44 $\pm$ 0.009\\
                   &               & $ 10.7 > \log \, (\Mstardot) \geq 10.46$ & 0.505 $\pm$ 0.007 &  0.475 $\pm$ 0.006 & 0.486 $\pm$ 0.01\\
\cline{2-6}
& \multirow{3}{*}{SF galaxies}&$ \log \, (\Mstardot)  \geq 11$ & 0.459 $\pm$ 0.01 &  0.541 $\pm$ 0.015 & 0.459 $\pm$ 0.011\\
 &                                 &$11 > \log \, (\Mstardot) \geq 10.4$ & 0.534 $\pm$ 0.007 &  0.543 $\pm$ 0.007 & 0.514 $\pm$ 0.012\\
  &                                &$10.4 > \log \, (\Mstardot) \geq 9.92$ & 0.578 $\pm$ 0.007 & 0.552 $\pm$ 0.007 & 0.549 $\pm$ 0.012\\
\hline
\multirow{2}{*}{Types} &\multirow{2}{*}{SF vs passive\tnotex{tnote:SF_q_mass}}& star-forming& 0.503 $\pm$ 0.007 & 0.491 $\pm$ 0.007 & 0.498 $\pm$ 0.007\\
 &                                 & passive& 0.462 $\pm$ 0.007 & 0.476 $\pm$ 0.007 & 0.467 $\pm$ 0.006\\                                  
\hline
\end{tabular*}
\begin{tablenotes}
     \item\label{tnote:panels} panels of Figure~\ref{Fig:PDF_dskel_G8Mpc_reshuff_dmatching}
     \item\label{tnote:median} medians of distributions as indicated in Figure~\ref{Fig:PDF_dskel_G8Mpc_reshuff_dmatching} by a vertical lines; errors are computed as in Table~\ref{tab:medians_dskel_dwall}  
     \item\label{tnote:before} as in Table~\ref{tab:medians_dskel_dwall} for $D_{\mathrm{skel}}/\left<D_\mathrm{z}\right>$                                     
     \item\label{tnote:after} reshuffling is done in bins of density computed at 8 Mpc (see the text for details)
     \item\label{tnote:dmatching} medians for the density-matched sample, where the density considered is computed at 8 Mpc 
     \item\label{tnote:SF_q_mass} only galaxies with stellar masses  $\log \, (\Mstardot) \geq 10.46$ are considered
    \end{tablenotes}
\end{threeparttable}
\end{table*}

\section{Discussion}
\label{sec:discussion}

Let us first discuss the observational  findings of the previous section in the  framework of existing work  (Section~\ref{sec:insight})
and then focus on a recent extension of anisotropic excursion set  which is developed in the companion paper  (Section~\ref{sec:EPS}).
The latter will allow us to explain why colour gradients prevail at fixed density.

\subsection{Cosmic web metric: expected impact on galaxy evolution}
\label{sec:insight}
%
%
In the  current framework for galaxy formation, in which galaxies  reside in extended dark matter halos, it is quite natural to split the environment into the \emph{local} environment, defined by the dark matter halo and the \emph{global} large-scale anisotropic  environment, encompassing the scale  beyond the halo's virial radius. 
The anisotropy of  the cosmic web is already a direct manifestation of the generic anisotropic nature of gravitational collapse on larger scales.
It provides the embedding in which dark halos and galaxies grow via accretion, which will act upon them via the combined effect of tides, the channeling of gas  along preferred directions and angular momentum advection onto forming galaxies.  

The observations and simulations presented in Sections~\ref{sec:mass_segregation}, \ref{sec:hagn} and~\ref{sec:density} provide a general support for this scenario. 
While rich clusters and massive groups are known to be environments which induce major galaxy transformations, the red fraction analysis presented in Section~\ref{subsec:red_frac} (Figure~\ref{Fig:red_frac}) reveals that the fraction of passive galaxies in the filaments starts to increase several Mpc away from the nodes and peaks in the nodes. This gradual increase suggests that some ``pre-processing" already happens before the galaxies reach the virial radius of massive halos and fall into groups or clusters \citep[e.g.][]{Porter2008, Martinez2016}. The above mentioned morphological transformation of elliptical-to-spiral ratio when getting closer to the filaments \citep[see also][]{Kuutma2017} can be interpreted as the result of mergers
transforming spirals into passive elliptical galaxies along the filaments when migrating towards nodes as suggested by theory and simulations \citep[][]{Codis2012,Dubois2014}. 
These findings show that  filamentary regions, corresponding to intermediate densities, are  important environments for galaxy transformation. This is also confirmed by the segregation found in Sections~\ref{subsec:fil} (Figures~\ref{Fig:PDF_dskel_masses} and~\ref{Fig:PDF_dskel_type}). More massive and/or passive galaxies are found closer to the core of filaments than their less massive and/or star-forming counterparts.
These differential mass gradients persist among the star forming population alone. In addition to mass segregation, star-forming galaxies show a gradual evolution in their star formation activity (see Figure~\ref{Fig:ssfr_u_r}). They are bluer than average at large distances from filaments ($D_{\rm skel} \gtrsim $ 5 Mpc), in a ``steady state" with no apparent evolution in star formation activity
at intermediate distances (0.5 $\le D_{\rm skel} \le$ 5 Mpc) and they show signs of decreased star formation efficiency near the core of the filaments ($ D_{\rm skel}\lesssim$ 0.5 Mpc).
These results are in line with the picture where on the one hand more massive/passive galaxies lay in the core of filaments and merge while drifting towards the nodes of the cosmic web. On the other hand, the low mass/star-forming galaxies tend to be preferentially located in the outskirts of filaments, a vorticity rich regions \citep{Laigle2015}, where galaxies acquire both their angular momentum (leading to a spin parallel to the filaments)  and their stellar mass  via essentially smooth accretion \citep[][also relying on \hagn]{Dubois2012a,Welker2017}. The steady state of star formation in these regions can reflect the right balance between the consumption and refuelling of the gas reservoir by the cold gas controlled by their surrounding filamentary structure \citep[as shown by][following \citealp{Pichon2011}, the outskirts of filaments are  the loci of most efficient helicoidal infall of cold gas]{Codis2015}. This may not be true anymore when galaxies fall in the core of the filaments. The decline of star formation activity can in part be due to the higher merger rate but also due to a quenching process such as strangulation, where the supply of cold gas is halted \citep[][]{Peng2015}. 
It could also find its origin in the cosmic web detachment \citep[][]{Aragoncalvo2016}, where the turbulent regions inside filaments prevent galaxies to stay connected to their filamentary flows and thus to replenish their gas reservoir.  
%
\begin{figure}
\includegraphics[width=1.\columnwidth]{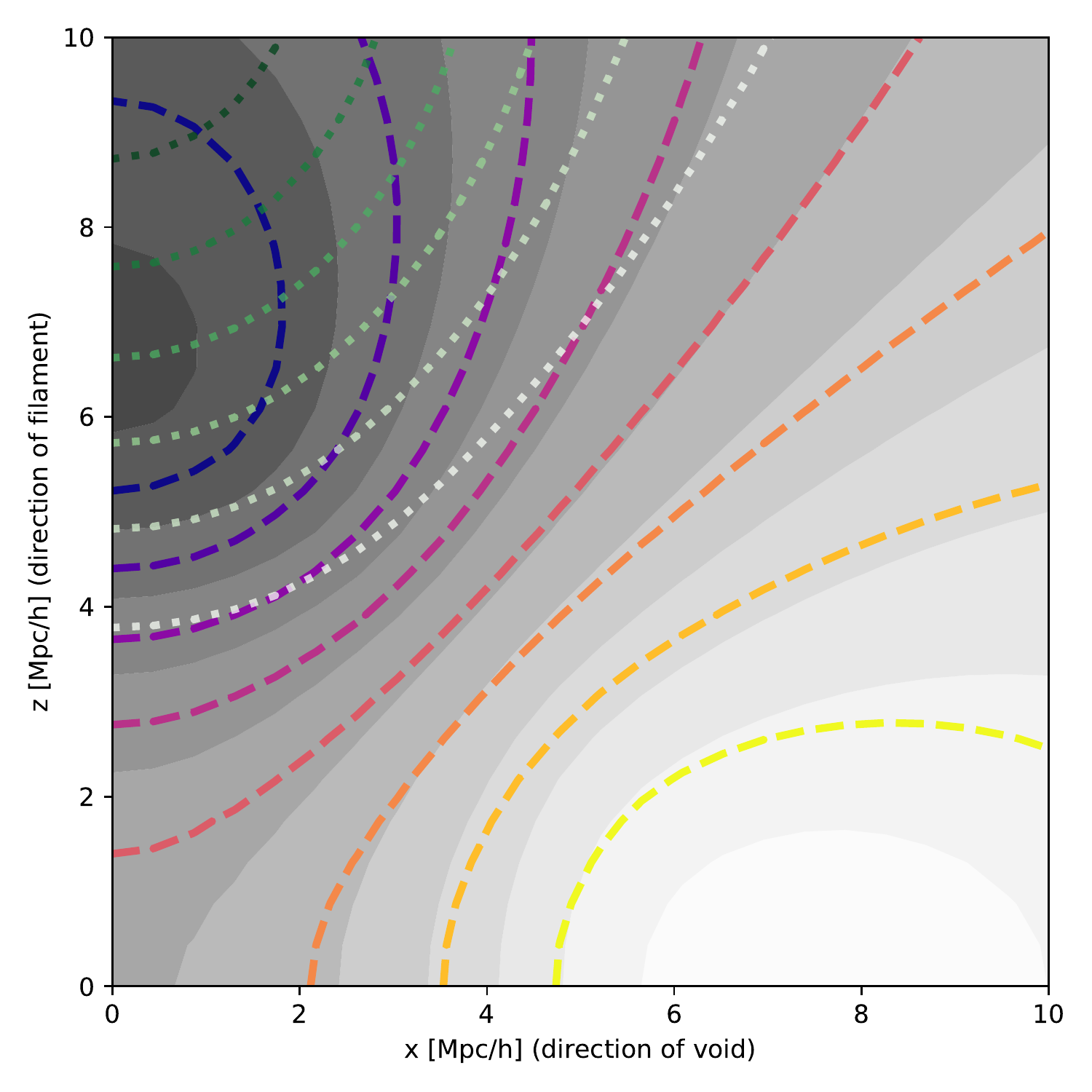}
\caption{Isocontours of constant typical  redshift $z=0$ mean density (filled contours), mass (dotted lines) and accretion rate (dashed lines)  in the frame of a filament (along the $Oz$ axis) in Lagrangian space (initial conditions) from low (light colours) to high values (dark colours). 
The saddle is at coordinate (0,0) while the induced peak and void are at coordinates (0,$\pm$7) and ($\pm$8,0) Mpc$/h$, respectively. 
As argued in the main text, this figure shows that the contours, hence the gradients of the three fields are not parallel (the contours cross). 
The choice of scale sets the units on the $x$ and $z$ axis (chosen  here to be 5 Mpc$/h$, while the mass and accretion rates are computed for 
a local smoothing of 0.5 Mpc$/h$).
At lower redshift/smaller scales, one expects the non-linear  convergence of the flow towards the filament to bring those contours together, aligning the gradients (see Figure~\ref{Fig:zeldovitch}). 
}
\label{Fig:Mnualpha}
\end{figure}

\subsection{Link with excursion set theory}
\label{sec:EPS}

The distinct transverse gradients found for mass, density and type or colour  may also be understood within the framework of 
conditional excursion set theory.
Qualitatively, the spatial variation of the (traceless part of the) tidal tensor in the vicinity of filaments will delay infall onto galaxies, which will impact  differentially galactic colour (at fixed mass), provided accretion can be reasonably converted  into star formation efficiency.

\subsubsection{Connecting gradients to constrained excursion set}
The companion paper \citep{biaspaper}
revisits excursion set theory subject to  conditioning the excursion to  the vicinity of  a filament.
In a nutshell, the main idea of  excursion set theory is to compute the statistical properties of the initial (over-)density  -- a stochastic variable -- enclosed within spheres of radius $R$, the scale which, through the spherical collapse model, can be related to the final mass of the object (should the density within the sphere pass the threshold for collapse).
Increasing  the radius of the sphere provides us with a proxy for ``evolution'' (larger sphere, larger mass, smaller variance, later formation time) 
{\it and} a measure of the impact of the environment (different sensitivity to tides for different, larger, spheres). 
The  expectations  associated to this stochastic variable can be re-computed subject to the tides imposed by larger-scale structures, which are best captured by the geometry of a filament-saddle point, $\cal S$, providing the  local natural ``metric''  for a filament \citep{Codis2015}. 
These large-scale tides will induce distinct weighting in the conditional PDF$(\delta,\partial_R \delta|{\cal S})$ for the over-density $\delta$, and its successive derivatives with respect to scale, $\partial_R \delta$ etc.  (so as to focus on collapsed  accreting regions). Indeed, the saddle will shift both the mean expectation of the PDFs but also  importantly their co-variances \citep[see][for details]{biaspaper}.
The derived  expected (dark matter) mean density $ \rho(r,\theta,\phi)$, Press-Schechter mass $ M(r,\theta,\phi)$ and typical accretion rate $ \dot M(r,\theta,\phi) $ then become explicit {\sl distinct}  functions of distance  $r$ and relative orientation  to the closest (oriented) saddle point. 
Within this model, it follows that the orientation of the mass, density and accretion rate gradients differ.
The misalignment arises because the various fields weight differently the constrained tides,  which will physically e.g. delay infall, and technically involve different moments of the aforementioned conditional PDF (see Appendix~\ref{sec:gradient} for more quantitative information on contour misalignment).
This is shown in  Figure~\ref{Fig:Mnualpha} which displays a  typical longitudinal cross section of those three maps in the frame of the saddle, with the filament along the $Oz$ axis, in Lagrangian space\footnote{This companion paper does not capture the strongly non-linear process of
dynamical friction of sub-clumps within dark matter halos, nor strong deviations from spherical collapse. We refer to  \cite{Hahn2009} which captures
the effect on satellite galaxies, and to \cite{Ludlow2014,Borzyszkowski2016,2016arXiv161103619C} which study the effect of the local shear on halos forming in filamentary structures.
This requires adopting a threshold for collapse that depends explicitly on the local shear.
The shear-dependent part of the critical density (and its derivative) correlates with the shear of the saddle, and introduces an additional anisotropic effect on top of the change of mean values and variances of density and slope.}.

This line of argument explains environmentally driven differential gradients, yet there is still a stretch to connect it to the observed gradients.
While there is no obvious consensus on the detailed effect of large-scale (dark matter) accretion onto the colour or star formation of galaxies at fixed mass and density, one can expect that the stronger the accretion, the stronger the AGN feedback, the stronger the quenching. Should this (reasonable) scaling hold true, the net effect in terms of gradients would be that colour gradients differ from mass and density ones. This is qualitatively consistent with the findings of this paper.

\subsubsection{Gradient alignments on smaller non linear scales}

The above presented Lagrangian theory clearly applies only on sufficiently large scales so that dynamical evolution has not driven the large scale flow too far from its initial configuration. On smaller scales, one would expect the same line of argument to operate in the frame set by the  saddle smoothed on the corresponding scale, but with one extra caveat: the increased level of non-linearity will have compressed the local filament transversally and stretched  it longitudinally,  following the generic kinetic flow measured in N-body simulation \citep[e.g.,][]{Sousbie2008a}, or predicted at the level of the Zel'dovich approximation \citep{Codis2015}.

\begin{figure}
\centering\includegraphics[width=0.8\columnwidth]{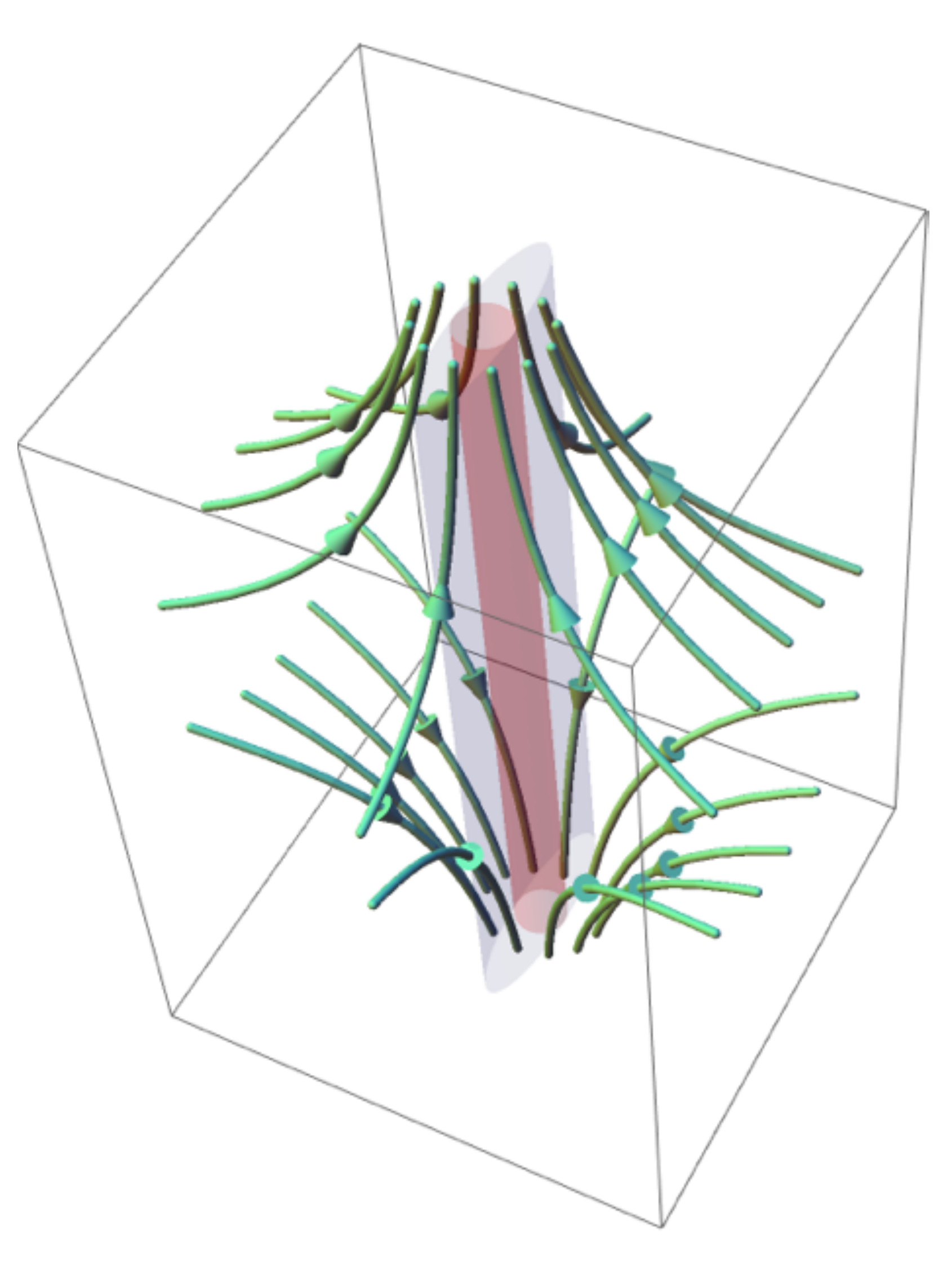}
\caption{Illustration of the Zel'dovich flow (green arrows) in the vicinity of a filament (red cylinder) embedded in a wall (purple flattened cylinder), with filament saddle at the centre. 
The non-linear evolution operating more strongly on smaller scales
will advect the contours presented in Figure~\ref{Fig:Mnualpha} along the green arrows, 
bringing them more parallel to each other. Consequently, at these smaller scales, the mass and accretion gradients do not differ significantly from 
the density gradients.  See \protect\cite{Codis2015} and \protect\cite{biaspaper} for more details.
}
\label{Fig:zeldovitch}
\end{figure}

Consequently, the contours of constant dark matter density $\rho$, typical dark halo mass $M$ and typical relative accretion rate  $\dot M/M$ in the frame of the saddle shown in Figure~\ref{Fig:Mnualpha} will be driven more parallel to each other, hence the difference in the orientation of the density, mass and accretion gradient will become smaller and smaller as one considers smaller scales, and/or more non linear dynamics (see Figure~\ref{Fig:zeldovitch}). As  colour gradient at fixed mass, and mass gradient at fixed density towards filaments originate from this initial misalignment, it should come as no surprise that as one probes smaller scales, such relative gradients disappear. When considering statistical expectations concerned with anisotropy (delayed accretion, acquisition of angular momentum, etc.), the net effect of past interactions should first be considered on the largest significant scale, beyond which the universe becomes isotropic. Conversely, the level of stochasticity should increase significantly on smaller scales, where one must account for e.g., the configuration of the last merger event, or the last fly-by.
Such a scenario is indeed supported by our findings in both GAMA and \hagn, presented in Appendices~\ref{sec:appendix_dtfe} and \ref{subsec:appendix_density_HzAGN}, Figures~\ref{Fig:PDF_dskel_dtfe_reshuff_dmatching} and \ref{Fig:PDF_dskel_reshuff_HzAGN}, respectively, whereby the use of the small scale density tracer does not allow to disentangle between the effects of the local density and that of cosmic web, suggesting that at such scale, they are closely correlated through the small-scale processes.

\subsubsection{Relationship to wall gradients}

When measured relative to the walls, galaxy properties are found to exhibit the same trends as for filaments, in that more massive and/or quiescent galaxies are found closer to the walls than their low mass and/or star-forming counterparts. This result is again in qualitative agreement with the idea of walls being, together with the filaments, the large-scale interference patterns of primordial fluctuations capable of inducing anisotropic boost in over-density together with the corresponding tides, and consequently imprinting their geometry in the measured properties of galaxies. The gradients measured for walls have the same origin as those inducing the differential gradients near the filament-type saddles, but are sourced by the geometry of the tides near the  wall-type saddles \citep[][Appendix~B]{Codis2015}. The main difference between the two saddles lies in the transverse curvatures, which is steeper for wall-type than for filament-type saddles (when considering the mean, eigenvolume weighted, eigenvalues of the curvature tensor with the relevant signatures) leading to weaker differences between the different gradients when considering walls.
This is consistent with the findings of Section~\ref{subsec:grad_wall}.

In closing, note that the (resp. Eulerian and Lagrangian) interpretations presented in Section~\ref{sec:insight} and~\ref{sec:EPS} are complementary, but fall short in explaining in details the origin of quenching. 
Nevertheless, in view of both observation and theory, the cosmic web metric appears  as a natural framework to understand  galaxy formation beyond stellar mass and local density.

\section{Summary and conclusions}
\label{sec:summary}

This paper studies the impact of the large-scale environment  on the properties of galaxies, such as their stellar mass, dust corrected $u - r$ colour and \ssfr. The discrete persistent structure extractor (DisPerSE) was used to identify the peaks, filaments and walls in the large-scale distribution of galaxies as captured by the GAMA survey. 
The principal findings are the following.
\begin{enumerate}
\item \textit{Mass segregation.} Galaxies are found to segregate by stellar mass, such that more massive galaxies are preferentially located closer to the cores of filaments than their lower mass counterparts. This mass segregation persists among the star-forming population.   
Similar mass gradients are seen with respect to walls in that galaxies with higher stellar mass tend to be found closer to the walls compared to galaxies with lower mass and persisting even when star-forming population of galaxies is considered alone.
\item \textit{Type/colour segregation.} Galaxies are found to segregate by type/colour, both with respect to filaments and walls, such that passive galaxies are preferentially located closer to the cores of filaments or walls than their star-forming counterparts. 
\item \textit{Red fractions.} The fraction of passive galaxies increases with both decreasing distance to the filament and to the node, i.e. at fixed distance to the node, the relative number of passive galaxies (with respect to the entire population) increases as the distance to the filament decreases and similarly, at a given distance to the filament, this number increases with decreasing distance to the node.    
\item \textit{Star formation activity.} Star-forming galaxies are found to carry an imprint of large-scale environment as well. Their dust corrected $u - r$ and \ssfr are found to be more enhanced and reduced, respectively, in the vicinity of the filaments compared to their outskirts.   
\item \textit{Consistency with cosmological  simulations.} All the found gradients are consistent with the analysis of the \hagn\, `full physics' hydrodynamical simulation. This  agreement suggests that what drives the gradients is captured by the implemented physics.  
\item \textit{Connection to excursion set theory.}  The origin of the distinct gradients can be qualitatively explained via conditional excursion set theory subject to filamentary tides \citep{biaspaper}. 
\end{enumerate}
This work has focused on filaments, nodes and in somewhat lesser details on walls. 
Similar observational results were recently reported at high redshift  by using the cosmic web filamentary structures in the VIPERS spectroscopic survey \citep{Malavasi2017} and when using  projected filaments in photometric redshift slices in the COSMOS field \citep{Laigle2017}.
These observations   are of intrinsic interest as a signature of galactic assembly; they also comfort theoretical expectations which point towards distinct gradients for colour, mass and density with respect to the cosmic web. The tides of the large-scale environment plays a significant specific role in the evolution of galaxies, and are  imprinted in their integrated physical properties,  which vary as a function of scale and distance to the different components of the cosmic web in a manner which is specific to each observable. 

These observations motivates a theory which eventually should integrate the anisotropy of the cosmic web as an essential ingredient  to i)  describe jointly  the dynamics and physics of galaxies, ii) explain galactic  morphological diversity, and iii) mitigate intrinsic alignment in upcoming lensing dark energy experiments, 
once a proper modelling of the mapping between galaxies and their halos (allowing e.g. to convert the DM accretion rate into colour of galaxy) becomes available.

Future large scale spectrographs on 8 meter class telescopes 
\citep[MOONS\footnote{Multi-Object Optical and Near-infrared Spectrograph};][PFS\footnote{Prime Focus Spectrograph; \href{http://pfs.ipmu.jp/}{http://pfs.ipmu.jp/}}; \citealp{Pfs}]{Moons1,Moons2} or space missions \citep[WFIRST\footnote{Wide-Field Infrared Survey Telescope; \href{http://wfirst.gsfc.nasa.gov}{http://wfirst.gsfc.nasa.gov}};][and Euclid\footnote{\href{http://sci.esa.int/euclid/}{http://sci.esa.int/euclid/}, \href{http://www.euclid-ec.org}{http://www.euclid-ec.org}}; \citealp{Euclid}, the deep survey for the latter]{Wfirst1,Wfirst2}
will extend the current analysis at higher redshift ($z \ge 1$) with similar samplings, allowing to explore the role of the environment near the peak of the cosmic star formation history, an epoch where the connectivity between the LSS and galaxies is expected to be even tighter, with ubiquitous cold streams. Tomography of the Lyman-$\alpha$ forest with PFS, MOONS, ELT-HARMONI \citep{Harmoni} tracing the intergalactic medium will make the study of the link between galaxies and this large scale gas reservoir possible (Laigle et al. in prep.).
 
\section*{Acknowledgments}
The authors thank the anonymous referee for suggestions and comments that helped to improve the presentation of the paper.
This research is carried out within the framework of the Spin(e) collaboration (ANR-13-BS05-0005, \href{http://cosmicorigin.org}{http://cosmicorigin.org}). 
We thank the members of this collaboration for numerous discussions. 
The \hagn \,simulation was post processed on the Horizon Cluster hosted by Institut d'Astrophysique de Paris. We thank S.~Rouberol for running it smoothly for us.
GAMA is a joint European-Australasian project based around a spectroscopic campaign using the Anglo-Australian Telescope. The GAMA input catalogue is based on data taken from the Sloan Digital Sky Survey and the UKIRT Infrared Deep Sky Survey. Complementary imaging of the GAMA regions is being obtained by a number of independent survey programmes including GALEX MIS, VST KiDS, VISTA VIKING, WISE, Herschel-ATLAS, GMRT and ASKAP providing UV to radio coverage. GAMA is funded by the STFC (UK), the ARC (Australia), the AAO, and the participating institutions. The GAMA website is \href{http://www.gama-survey.org/}{http://www.gama-survey.org/}.
The VISTA VIKING data used in this paper is based on observations made with ESO Telescopes at the La Silla Paranal Observatory under pro- gramme ID 179.A-2004.
CC acknowledges support through the ILP PhD thesis fellowship. CL is supported by a Beecroft Fellowship.

\bibliographystyle{mnras}
\bibliography{cw_gama}
\appendix

\section{Matching technique}
\label{sec:appendix_matching}

\subsection{Mass matching}
\label{subsec:appendix_mmatching}

First the mass distributions of the two populations are cut so that they cover the same stellar mass range, i.e. they have the same minimum and maximum value of stellar mass. Then, in each stellar mass bin, the population with lower number of galaxies is taken as the reference sample and $N_{\mathrm{match}}$ samples of galaxies are extracted in the other population, such that their mass distribution is the same as the one of the reference sample. In practice, for each galaxy in the reference sample, the corresponding galaxy of the larger sample is sought among galaxies whose mass difference with respect to the reference mass is smaller than $\Delta M_{\star}$ in logarithmic space. 
If there is no galaxy in larger sample satisfying this condition, the galaxy of the reference sample is removed from the analysis. 
In each of $N_{\mathrm{match}}$ samples every galaxy of the larger sample is considered only once, however repetitions are allowed across all samples. 
By construction, after applying this procedure, one ends up with $N_{\mathrm{match}}$ samples consisting of the same number of star-forming and passive galaxies and having very similar stellar mass distributions. 

If not stated differently, 20 mass-matched samples are typically constructed using ten equipopulated stellar mass bins for each and choosing a value of 0.1 for  $\Delta M_{\star}$ parameter. Varying the values of  $N_{\mathrm{match}}$, $\Delta M_{\star}$ and the number of stellar mass bins within the reasonable range does not alter our conclusions.

\subsection{Density matching}
\label{subsec:appendix_dmatching}

This Appendix provides details on the density matching procedure. First, let us describe how the mass-density matched samples are constructed. The galaxy sample is first divided into three logarithmic stellar mass bins for which the density matched samples are to be constructed. In each of the 10 equipopulated logarithmic over-density $1+\delta$ bins the reference sample is identified as that of the previously constructed  stellar-mass subsamples with the lowest number of galaxies.
Next, for each galaxy in the reference sample, a galaxy is randomly chosen from each of two stellar mass bins having the over-density closest to the galaxy in the reference sample. In practice, the nominal absolute difference in the $\log (\Mstardot)$ values used to match galaxies is 0.1. If no suitable galaxy is found in at least one of the two stellar mass bins, the galaxy of the reference sample is removed from the analysis. This procedure is repeated ten times, ending up with ten samples of galaxies having the same over-density distributions in three different stellar mass bins . 

Similarly, to construct type-density matched samples, the entire galaxy sample is first divided into the subsamples of star-forming and passive galaxies. Then, in each of the ten equipopulated logarithmic over-density $1+\delta$ bins, the reference sample (sample of passive or star-forming galaxies) is identified as the one having the lowest number of galaxies. We continue by randomly choosing a galaxy from the larger sample with an over-density and stellar mass close to that of the galaxy from the reference sample. In practice, we pair galaxies for which the distance in the two-parameter logarithmic space, defined by the stellar mass and the over-density, is minimal and smaller than 0.1. The procedure is again repeated ten times in order to construct ten samples of star-forming and passive galaxies having their mass and density distributions close to each other.

\begin{figure*}
\includegraphics[width=0.7\textwidth]{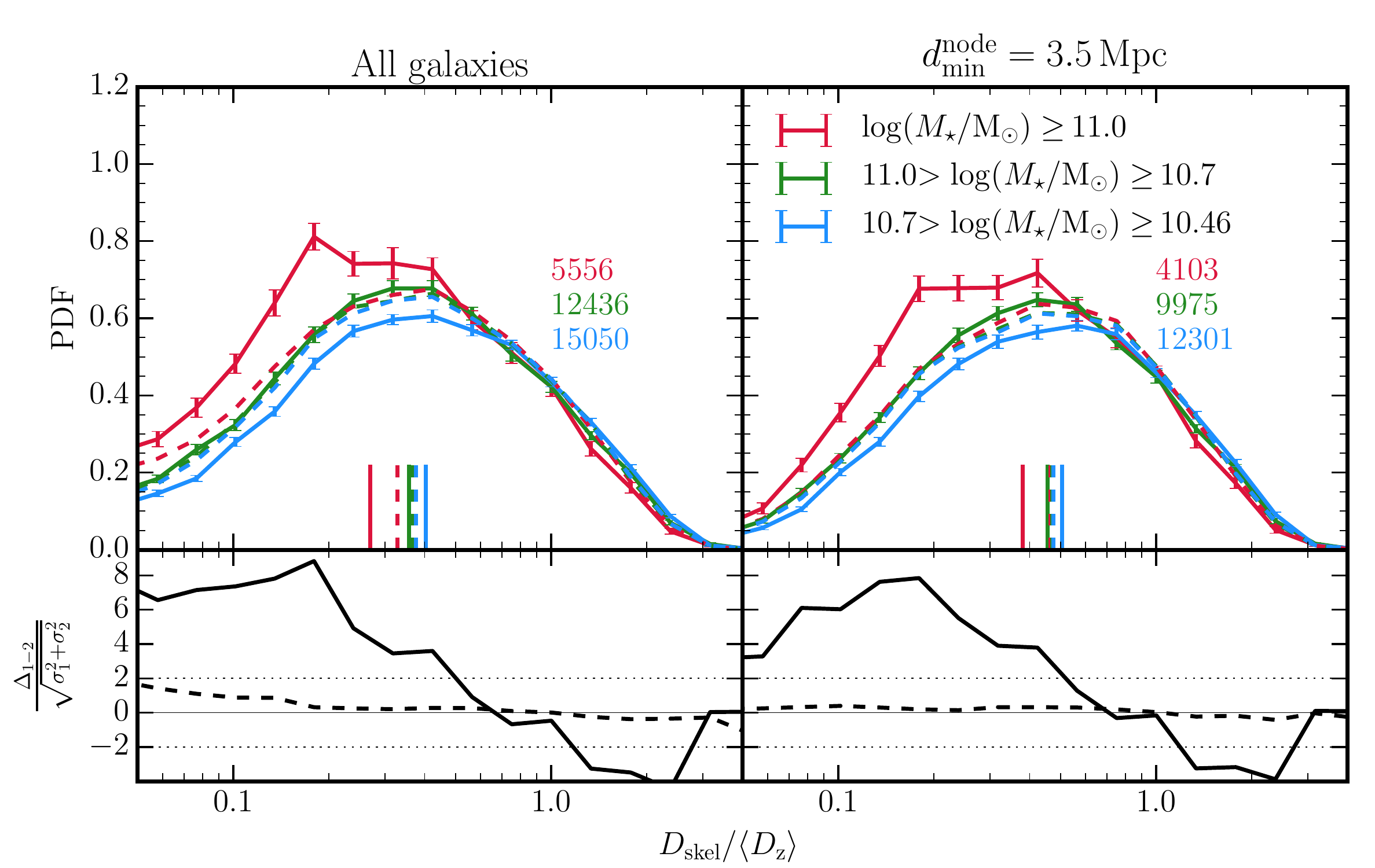}
\caption{\textit{Top row:} Differential distributions of the normalised distances to the nearest filament, $D_{\mathrm{skel}}$. 
The solid lines show mass gradients for all galaxies (left panel) and after removing galaxies with distances to the node smaller than 3.5 Mpc (right panel). The dashed lines illustrate mass gradients after the reshuffling of $D_{\mathrm{skel}}$ of galaxies in bins of distances to the node $D_{\mathrm{node}}$. The vertical lines indicate the medians of the distributions and their values together with associated errors are listed in Table~\ref{tab:mass_gradients_dskel_dnode}. 
The reshuffling method cancels mass gradients towards filaments once galaxies at distances closer than 3.5 Mpc from nodes are removed. 
\textit{Bottom row:} Residuals in units of $\sigma$ between the two most extreme mass bins ($\log(\Mstardot) \geq 11.0$ and $10.7 > \log (\Mstardot) \geq 10.46$) before (solid lines) and after (dashed lines) the reshuffling.}
\label{Fig:appendix_dskel_dnode}
\end{figure*}

\section{The impact of cosmic boundaries} 
\label{sec:appendix_boundaries}

It was stated in Sections~\ref{subsec:grad_skel} and \ref{subsec:grad_wall} that the measured gradients towards filaments (Figures~\ref{Fig:PDF_dskel_masses} and \ref{Fig:PDF_dskel_type}) and walls (Figures~\ref{Fig:PDF_dwall_masses} and \ref{Fig:PDF_dwall_type}) are not simply due to gradients towards nodes in the former and due to gradients towards nodes and filaments in the latter case. 
 This Appendix presents the performed tests that allowed us to reach such a conclusion. 

\subsection{Gradients towards filaments}
\label{subsec:appendix_filaments}

Let us start by considering the gradients towards filaments. In order to probe these gradients without being substantially contaminated by the contribution from nodes, galaxies that are closer to nodes than 3.5 Mpc are removed from the analysis. The choice of this distance $d^{\mathrm{node}}_{\mathrm{min}}$ is motivated by the compromise between eliminating the most of the gradient coming from nodes while keeping enough objects to have a statistically significant sample. 
Note that the distance of 3.5 Mpc is greater than the typical size of groups, which is $\sim$ 1.5 Mpc in the redshift range considered in this work, measured as a median (or mean) projected group radius.
The value of median (and mean) is insensitive to the definition of the group radius \citep[see][for various definitions considered]{Robotham2011}. 
In Figure~\ref{Fig:appendix_dskel_dnode}, the solid lines show the mass gradients towards filaments for the entire  sample (left panel) on the one hand and after excluding galaxies with distances to the node $D_{\mathrm{node}} \leq 3.5$ Mpc (right panel). 

The contribution of nodes to  mass gradients towards filaments is measured by randomising distances to the filament, $D_{\mathrm{skel}}$, in bins of distances to the node, $D_{\mathrm{node}}$. By construction,  gradients towards nodes are preserved.  
20 samples are constructed in each of which this reshuffling method is applied in 20 equipopulated logarithmic bins.  
As shown by the dashed lines in Figure~\ref{Fig:appendix_dskel_dnode} and values of medians listed in Table~\ref{tab:mass_gradients_dskel_dnode}, the reshuffling cancels the gradients towards filaments for $d^{\mathrm{node}}_{\mathrm{min}}$ = 3.5 Mpc.

In addition, following \cite{Laigle2017}, it can be shown that in the regions sufficiently faraway from nodes, gradients towards nodes and those towards filaments are independent. 
It was checked that the mass gradients towards nodes, present for the entire galaxy sample, are substantially reduced once galaxies for which distances to the node $D_{\mathrm{node}} \leq 3.5$ Mpc are excluded.
This time, the distances to the node, $D_{\mathrm{node}}$, were randomised in bins of distances to the filament, $D_{\mathrm{skel}}$, i.e. by construction, gradients towards filaments were preserved. Again, 20 samples were constructed using 20 equipopulated logarithmic bins. After reshuffling, weak gradients at the level of at most 1$\sigma$ are still present, but  note that additional increase in $d^{\mathrm{node}} _{\mathrm{min}}$ does not reduce them further.

This analysis allows us to conclude that by removing from our sample galaxies that are closer to nodes than 3.5 Mpc, the impact of nodes to the measured gradients towards filaments is minimised, and even if weak gradients towards nodes still exist, these are independent of gradients towards filaments, i.e. gradients towards filaments and gradients towards nodes can be disentangled. 

Let us finish this section with two remarks. First, note that distances to the node considered here are 3D euclidian distances. Curvilinear distances along the filaments could have been used instead (as illustrated in Figure~\ref{Fig:sketch_filament}). This alternative choice of the distance does not alter our conclusions.
Secondly, instead of using distances to the node $D_{\mathrm{node}}$, one could have considered distances normalised by the redshift-dependent mean inter-galaxy separation, $D_{\mathrm{node}}/\langle D_{\mathrm{z}} \rangle$. These two approaches give consistent results not only qualitatively, but also quantitatively. 

\subsection{Gradients towards walls}
\label{subsec:appendix_walls}

\begin{figure*}
\includegraphics[width=0.7\textwidth]{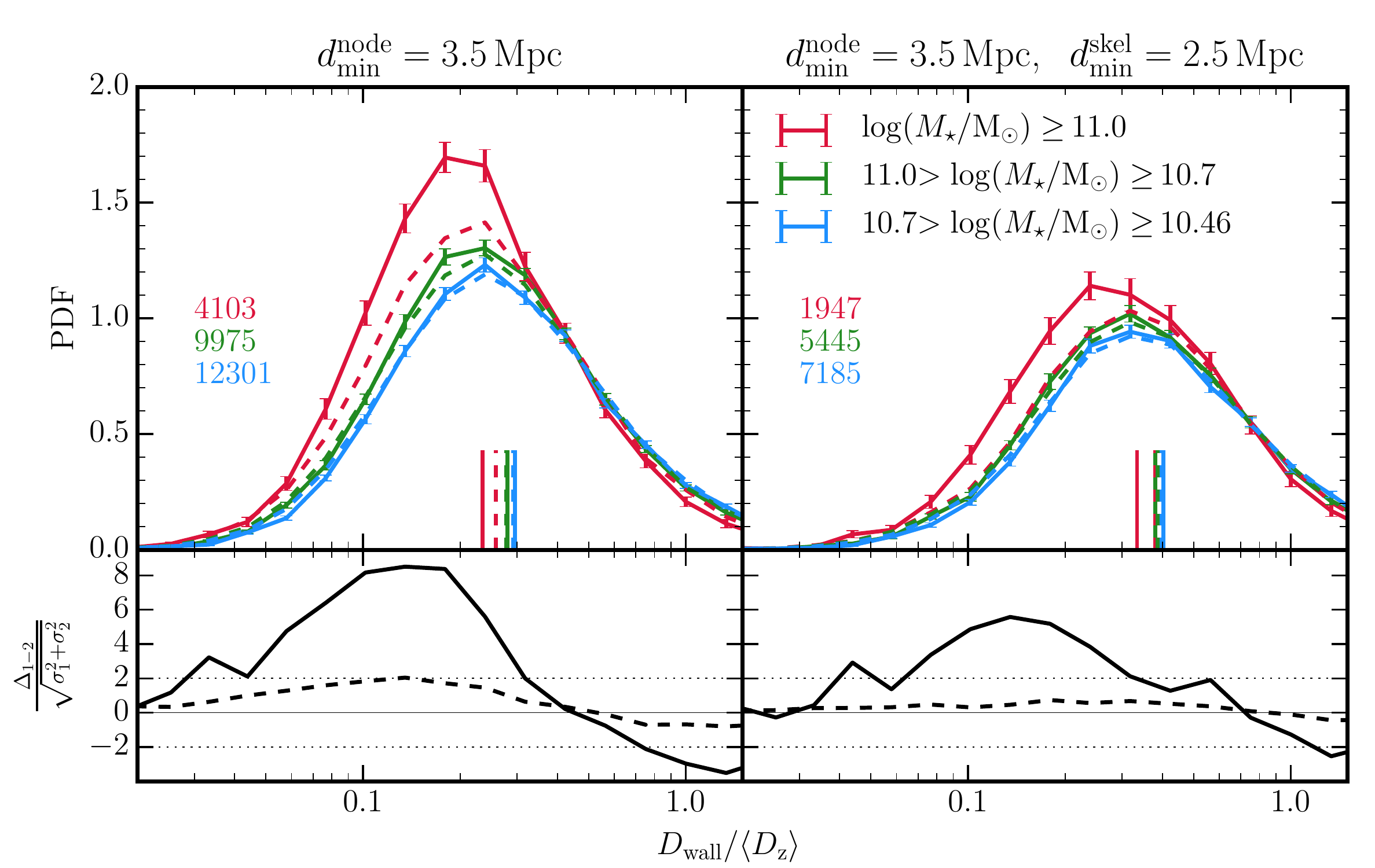}
\caption{\textit{Top row:} Differential distributions of the normalised distances to the nearest wall, $D_{\mathrm{wall}}$. 
The solid lines show mass gradients after removing galaxies with distances to the node smaller than 3.5 Mpc (left) and after applying an additional criteria on the distance to the filament, such that galaxies with distances to the filament smaller than 2.5 Mpc (right) are removed. The dashed lines illustrate mass gradient after reshuffling of $D_{\mathrm{skel}}$ of galaxies in bins of distances to the node $D_{\mathrm{node}}$. As shown on the right panel, these are almost completely cancelled after removing sufficiently large regions around nodes and filaments.
The vertical lines indicate the medians of the distributions and their values together with associated errors are listed in Table~\ref{tab:mass_gradients_dwall_dskel}. 
\textit{Bottom row:} Residuals in units of $\sigma$ as in Figure~\ref{Fig:appendix_dskel_dnode}.}
\label{Fig:appendix_dwall_dskel}
\end{figure*}

As with filaments, when measuring the gradients towards walls, one should investigate whether the gradient is not dominated by other component of the  environments. As filaments are regions where walls intersect,  these represent  on top of nodes 
an additional source of contamination for the measured gradients  towards walls.
Figure~\ref{Fig:appendix_dwall_dskel} shows the mass gradients towards walls for the galaxy sample outside the zone of influence of nodes parametrised by $d^{\mathrm{node}}_{\mathrm{min}}$ = 3.5 Mpc (left panel) and after applying an additional criterion by excluding galaxies with distances to the closest filament 
$D_{\mathrm{skel}} \leq d^{\mathrm{skel}}_{\mathrm{min}}$ with $d^{\mathrm{skel}}_{\mathrm{min}}$ = 2.5 Mpc (right panel). 
The contribution of filaments to the mass gradients towards walls is measured by randomising distances to the wall, $D_{\mathrm{wall}}$, in bins of distances to the filament, $D_{\mathrm{skel}}$. By construction, the gradients towards filaments are preserved. 
Here 20 samples are constructed in each of which the reshuffling method is applied in 20 equipopulated logarithmic bins.  
As shown by the dashed lines in Figure~\ref{Fig:appendix_dwall_dskel} and values of medians listed in Table~\ref{tab:mass_gradients_dwall_dskel}, the reshuffling cancels the gradients towards walls for $d^{\mathrm{skel}}_{\mathrm{min}}$ = 2.5 Mpc.

Following the method used in Appendix~\ref{subsec:appendix_filaments} it was verified (but not shown here) that the mass gradients towards filaments after randomisation of the distances $D_{\mathrm{skel}}$ in bins of distances to the nearest wall $D_{\mathrm{wall}}$ are substantially reduced. Only a very weak mass gradient (at  a 1$\sigma$ level at most) is detected after randomisation even for $d^{\mathrm{skel}}_{\mathrm{min}}$ = 2.5 Mpc. Similarly to what was found in Section~\ref{subsec:appendix_filaments},  increasing  this parameter does not  induce any substantial  reduction of the gradient.  Thus  this distance was chosen as the limit for the exclusion region around filaments.

\section{Small scale density-cosmic web relation}
\label{sec:appendix_dtfe}

\begin{figure*}
    \centering
    \begin{subfigure}{\textwidth}
        \centering
        \includegraphics[width=\textwidth]{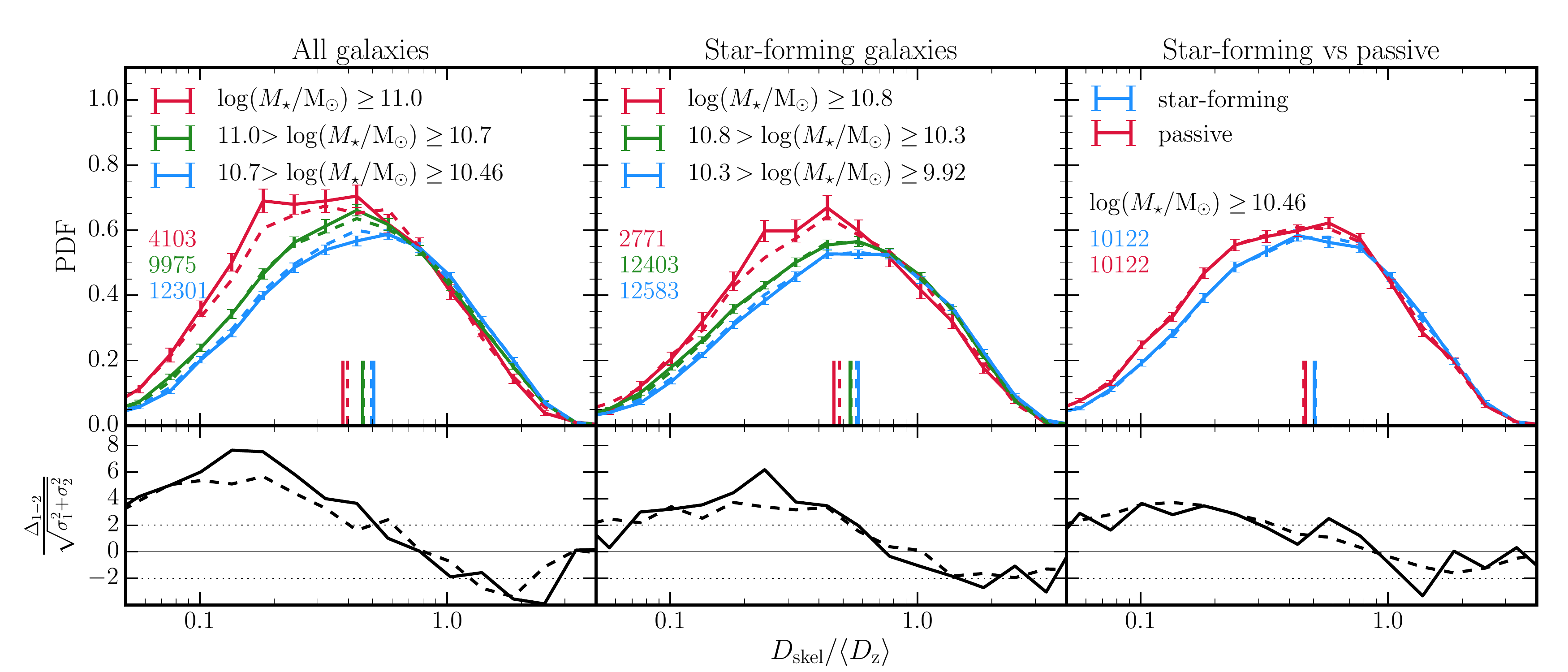}
        \caption{Reshuffling.}
    \end{subfigure}%
    \\
    \begin{subfigure}{\textwidth}
        \centering
        \includegraphics[width=\textwidth]{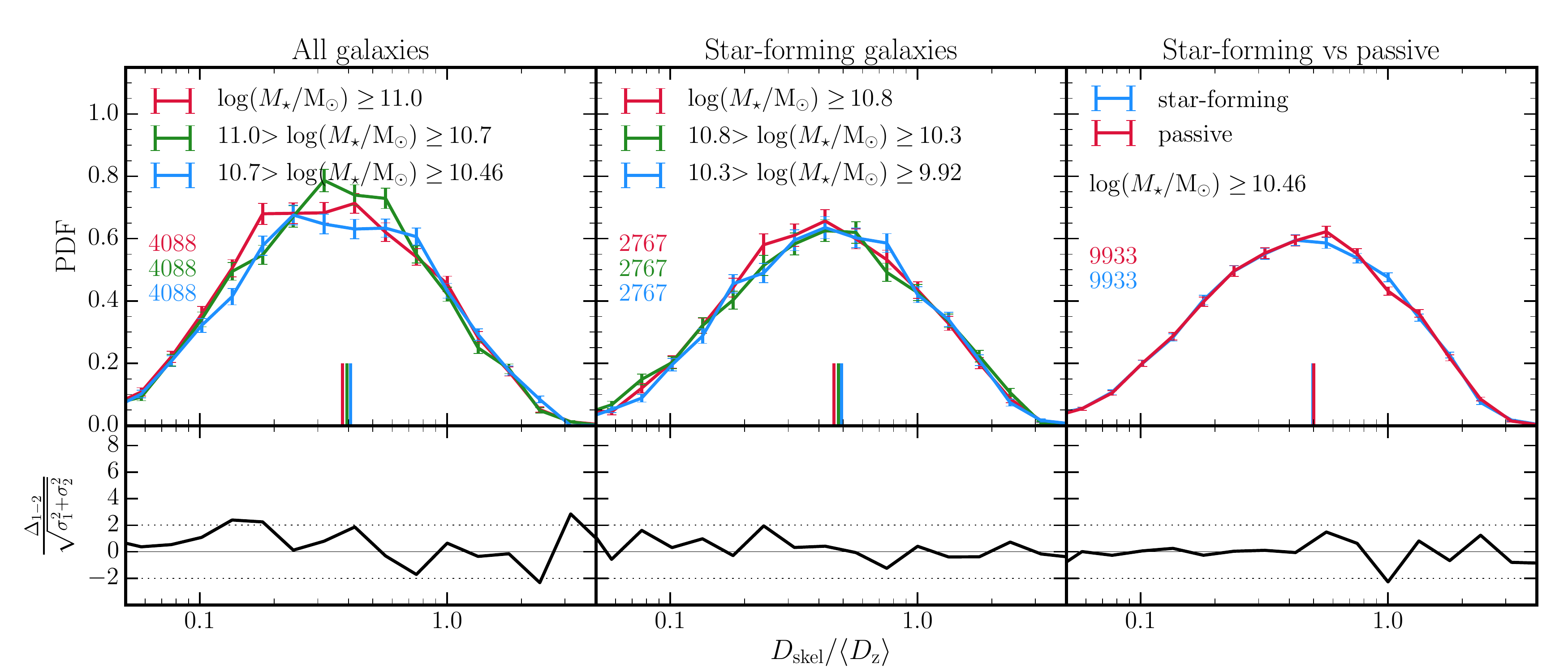}
        \caption{Density matching.}
    \end{subfigure}
    \caption{\textit{Top rows:} As in Figure~\ref{Fig:PDF_dskel_G8Mpc_reshuff_dmatching}, but using the DTFE density for both methods, reshuffling (Figure a) and density matching (Figure b).
The numerical values of medians, shown as a vertical lines, are listed in Table~\ref{tab:medians_dskel_dtfe_reshuff_dmatching}. 
When the small-scale density, DTFE in this case, is used in the reshuffling method, the randomised (dashed lines) and original signal (solid lines) are nearly identical. Similarly, all gradients are almost completely erased, as expected.
 \textit{Bottom rows:} Residuals in unit of $\sigma$ as in Figure~\ref{Fig:PDF_dskel_G8Mpc_reshuff_dmatching}.
 }
    \label{Fig:PDF_dskel_dtfe_reshuff_dmatching}
\end{figure*}

In this Appendix, the impact of the small-scale density estimator on the mass and type/colour gradients is presented. The density used here is DTFE, i.e. the density computed at the smallest possible scale\footnote{There is no specific scale associated to the DTFE: it is a local adaptive method which determines the density at each point  while preserving its multi-scale character.}.  As in the Section~\ref{sec:density}, the two methods, the reshuffling and density-matching, are applied.

Figure~\ref{Fig:PDF_dskel_dtfe_reshuff_dmatching} shows the differential distributions of the distances to the nearest filament, $D_{\mathrm{skel}}$ (normalised by $\langle D_{\mathrm{z}} \rangle$, for the same selections as in Figure~\ref{Fig:PDF_dskel_G8Mpc_reshuff_dmatching}. 
The contribution of the nodes to the measured signal is minimised, by removing from the analysis galaxies located closer to a node than 3.5 Mpc. Star-forming and passive galaxies have been matched in mass, as described in Appendix~\ref{subsec:appendix_mmatching}. The vertical lines indicate the medians of the distributions, whose values together with the error bars are listed in Table~\ref{tab:medians_dskel_dtfe_reshuff_dmatching}. 

In Figure (a), the mass and type gradients are shown before (solid lines, as in~\ref{Fig:PDF_dskel_G8Mpc_reshuff_dmatching}) and after (dashed lines) applying the reshuffling of galaxies in the bins of over-density $1 + \delta$, where the number density corresponds to the DTFE density. 
The result conforms to the expectations. The reshuffling does not remove the observed mass and type/colour gradients, i.e. the distributions before and after the reshuffling are almost identical, suggesting that at the small scale, traced by DTFE, the density and cosmic web are closely correlated through the small-scale processes.   

Figure (b) illustrates the PDFs for samples that have been matched in over-density $1 + \delta$, as described in Appendix~\ref{subsec:appendix_dmatching}, where the density considered is DTFE. The density-matching technique yields qualitatively similar result than the above used reshuffling in that almost no mass and type gradients are detected when galaxies matched in the DTFE density.

Qualitatively same results are obtained for both methods when applied to the measurements of gradients with respect to the walls (not shown) .

\section{The HORIZON-AGN simulation}
\label{sec:appendix_hagn}
 
This Appendix is dedicated to presenting the large-scale cosmological hydrodynamical simulation \hagn~\citep{Dubois2014}.  
First, some of the main features of the simulation are briefly summarised. The reshuffling method is then implemented on the simulation, as defined in Section~\ref{sec:density}, and shown to yield qualitatively similar results  to those obtained in GAMA for both large- and small-scale density tracers.  
 
\subsection{Simulation summary} 
 \label{subsec:appendix_hagn_summary}
 
The detailed description of the \hagn\,simulation\footnote{ \href{http://www.horizon-simulation.org}{http://www.horizon-simulation.org}} can be found in~\cite{Dubois2014}, here only its brief summary is given. 
The cosmological parameters used in the simulation correspond to the $\Lambda$CDM cosmology with total matter density $\Omega_{\rm  m}=0.272$, dark energy density $\Omega_\Lambda=0.728$, amplitude of the matter power spectrum $\sigma_8=0.81$, baryon density $\Omega_{\rm  b}=0.045$, Hubble constant $H_0=70.4 \, \rm km\,s^{-1}\,Mpc^{-1}$, and $n_s=0.967$ compatible with the WMAP-7 data~\citep{Komatsu2011}. 

The simulation was run with the Adaptive Mesh Refinement code RAMSES~\citep{Teyssier2002} in a box of length $L_{\rm box} = 100\, h^{-1}\, \rm Mpc$ containing $1024^3$ dark matter (DM) particles, with a DM mass resolution of $M_{\rm  DM, res}=8\times10^7 \, \rm M_\odot$, and initial gas resolution of $M_{\rm gas,res}=1\times 10^7 \, \rm M_\odot$. 
 
The collisionless DM and stellar components are evolved using a particle-mesh solver. The dynamics of the gaseous component are computed by solving Euler equations on the
adaptive grid using a second-order unsplit Godunov scheme.

 The refinement is done in a quasi-Lagrangian manner starting from the initial coarse grid down to $\Delta x=1$ proper kpc (seven levels of refinement) as follows: each AMR cell is refined if the number of DM particles in a cell is more than eight, or if the total baryonic mass in a cell is eight times the initial DM mass resolution. This results in a typical number of $7\times 10^9$ gas resolution elements (leaf cells) in the \hagn\, simulation at $z=0$.
  
Heating of the gas from a uniform UV background takes place after redshift $z_{\rm  reion} = 10$ following~\cite{Haardt1996}. Gas is allowed cool down to $10^4\, \rm K$ through H and He collisions with a contribution from metals using a~\cite{Sutherland1993} model. 

The conversion of gas into stars occurs in regions with gas density exceeding $\rho_0=0.1\, \rm H\, cm^{-3}$ following the~\cite{Schmidt1959} relation of the form $\dot \rho_*= \epsilon_* {\rho_{\rm g} / t_{\rm  ff}}$, where $\dot \rho_*$ is the star formation rate mass density, $\rho_{\rm g}$ the gas mass density, $\epsilon_*=0.02$ the constant star formation efficiency, and $t_{\rm  ff}$ the local free-fall time of the gas.

Feedback from stellar winds, supernovae type Ia and type II are included into the simulation with mass, energy and metal release. \hagn\,simulation takes also into account the formation of black holes (BHs) that can grow by gas accretion at a Bondi-Hoyle-Lyttleton rate capped at  the Eddington accretion rate when they form a tight enough binary.
The AGN feedback is a combination of two different modes (the so-called quasar and radio mode) in which
BHs release energy in the form of heating or jet when the accretion rate is respectively above and below one per cent of Eddington, with efficiencies tuned to match the BH-galaxy scaling relations at $z=0$~\citep[see][for details]{Dubois2012}.

Galaxies are identified using the updated method~\citep{Tweed2009} of the AdaptaHOP halo finder~\citep{Aubert2004} directly operating on the distribution of stellar particles. Only galactic structures with a minimum of $N_{\rm  min}=100$ stellar particles are considered, which typically selects objects with masses larger than  $2 \times 10^8 \, \rm M_\odot$.

\subsection{Density reshuffling}
\label{subsec:appendix_density_HzAGN}
\balance
Let us finally present the impact of the reshuffling method, as defined in Section~\ref{sec:density}, and the choice of the density tracer in the \hagn\,simulation.   

\begin{figure*}
    \centering
    \begin{subfigure}{\textwidth}
        \centering
        \includegraphics[width=0.75\textwidth]{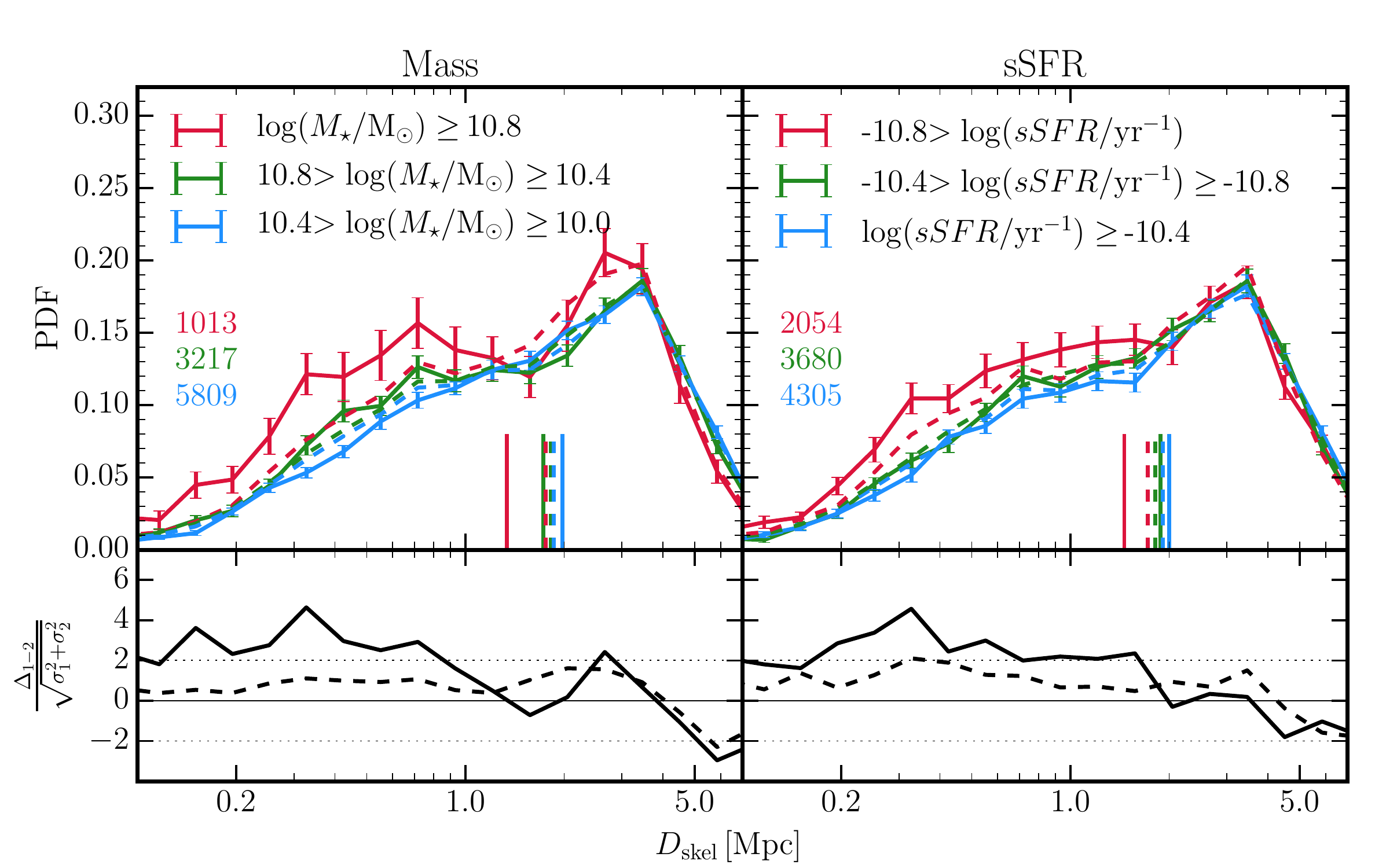}
        \caption{Reshuffling using the density computed in the Gaussian kernel at the scale of 5 Mpc.}
    \end{subfigure}%
    \\
    \begin{subfigure}{\textwidth}
        \centering
        \includegraphics[width=0.75\textwidth]{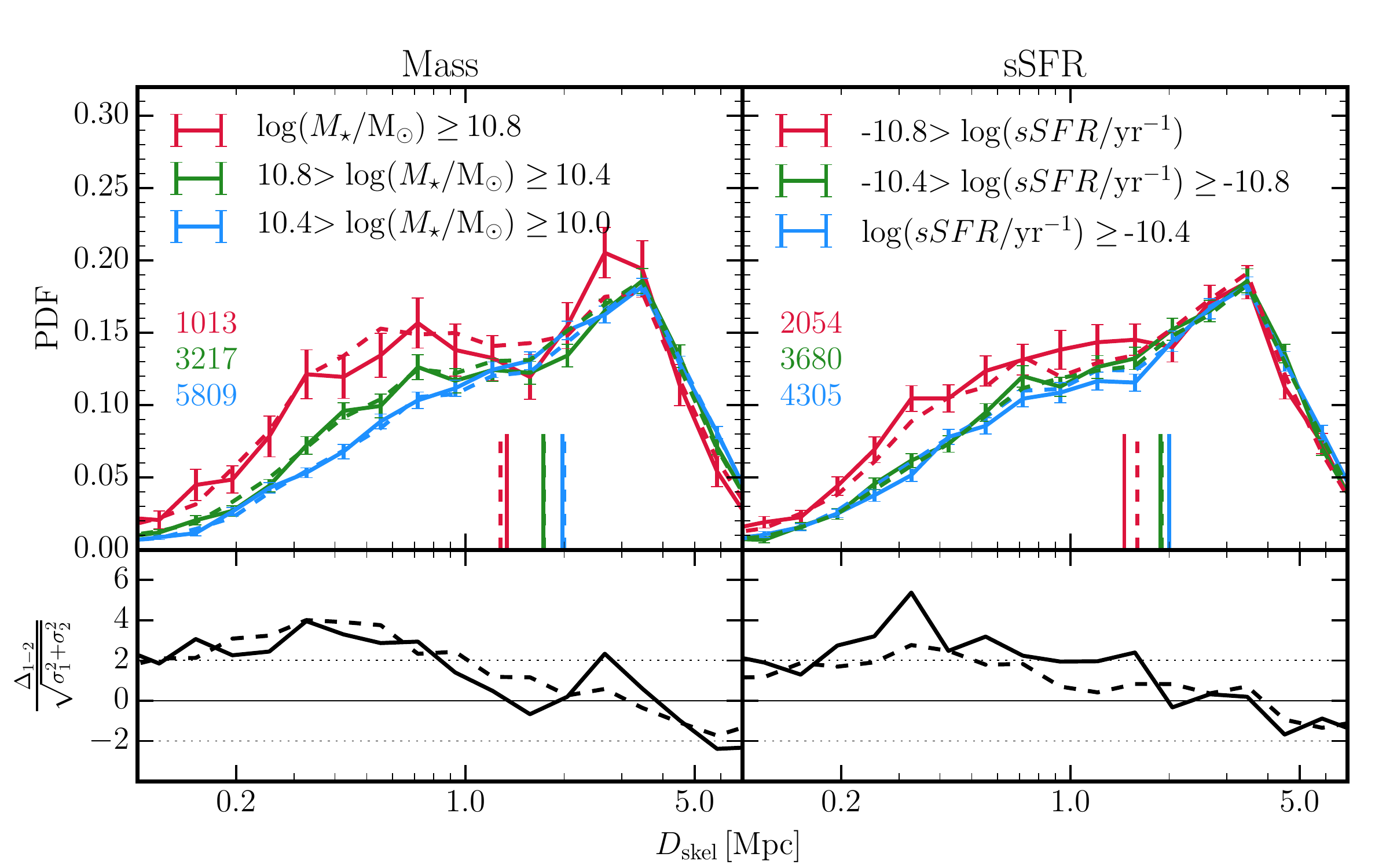}
        \caption{Reshuffling using the DTFE density.}
    \end{subfigure}
    \caption{\textit{Top rows:} As in Figure~\ref{Fig:PDF_dskel_dwall_HzAGN} for the distances to the nearest filament, $D_{\mathrm{skel}}$. The contribution of the nodes is minimised by removing galaxies located within 3.5 Mpc around them from the analysis. The dashed lines correspond to the distributions after the application of the reshuffling method using two different density tracers, a large (Figure a) and small-scale (Figure b) estimators. The numerical values of medians, shown as a vertical lines, are listed in Table~\ref{tab:medians_dskel_dens_reshuff_hagn}. In qualitative agreement with the results obtained with the observed data, in order to cancel the gradients, density at sufficiently large scale has to be considered. This corresponds to 5 Mpc in the \hagn\,simulation, representing $\sim$ 1.5$\times$ mean inter-galaxy separation, again in agreement with the value found in observations.   
 \textit{Bottom rows:} As in Figure~\ref{Fig:PDF_dskel_dwall_HzAGN} before (solid lines) and after (dashed lines) the reshuffling.
 }
    \label{Fig:PDF_dskel_reshuff_HzAGN}
\end{figure*}

Figure~\ref{Fig:PDF_dskel_reshuff_HzAGN} 
illustrates that the result of reshuffling depends on the scale at which the density is computed. As expected, when using the small-scale density tracer, such as e.g. the DTFE density (Figure a), both mass and \ssfr gradients are almost unchanged, while on sufficiently large scales, the gradients tend to cancel out (Figure b). The numerical value of the scale at which this happens is $\sim$ 5 Mpc. This is again in a qualitative agreement with the scale required in the GAMA survey, corresponding to the $\sim$ 1.5 $\times$ mean inter-galaxy separation.
  

\section{Gradient misalignments}
\label{sec:gradient}

In the context of conditional excursion set theory subject to a saddle $\S$ at some finite distance $(r,\theta,\phi)$ from a forming halo, let us consider the Hessian of the potential, 
$  q_{ij} \equiv \partial^2 \psi/\partial r_i\partial r_j$,  smoothed on the saddle scale $\Rs$ and normalized so that $\mean{\tr^2(\mathbf q)}=1$.  
The anisotropic shear is given by the traceless part $\bar q_{ij} \equiv q_{ij}-\delta_{ij}\tr{\,\mathbf q}/3$, which deforms the region by slowing down or accelerating the collapse along each axis. At finite separation, this traceless shear modifies in an anisotropic way the statistics of  the smooth mean density (and of its derivative with respect to scale).
The variations are modulated by $\mathcal{Q}=\sum_{i,j}{\hat{r}_i\bar q_{ij}\hat{r}_j}$, with $\hat r_i=r_i/r$, i.e. by the relative orientation of the separation vector, $\mathbf r$, in the frame set by the tidal tensor of the saddle. This extra degree of freedom, $\mathcal{Q}(\theta,\phi)$, provides a supplementary vector space, beyond the radial direction, over which to project the gradients, with statistical weight depending on each specific observable (mass, accretion rate, etc.). These quantities have thus potentially different iso-surfaces from each other and from the local mean density, a genuine signature of the impact of the traceless part of the tidal tensor.  Indeed, for each observable, the conditioning on $\mathcal{S}$ introduces a further  dependence on the geometry of the environment (the height  of the saddle and its anisotropic shear $\bar q_{ij}$) and on the position $\mathbf r$ of the halo with respect to the saddle point. This dependence arises because the saddle point condition modifies  the mean and variance of the stochastic process $(\delta,\partial_R \delta)$ -- the height and slope of the excursion set trajectories -- in a position-dependent way, making it more or less likely to form halos of given mass and assembly history within the environment set by $\cal S$. The expectation of the process becomes anisotropic through ${\cal Q}$, and both mean and variance acquire distinct radial dependence through the the relevant correlation functions $\xi_{\alpha\beta}$ defined below in Equation~\eqref{eq:xi_xi}.

For instance, considering the typical mass, $M_\star$, and accretion rate, ${\dot M}_\star$, at scale $R$, straightforward trigonometry shows that cross product of their gradients reads
\begin{equation}
  \left(\frac{\partial  \dot M_\star}{\partial r} \frac{\partial
      M_\star }{\partial {\cal Q}}-\frac{\partial  \dot M_\star}{\partial {\cal Q}} \frac{\partial M_\star }{\partial r}\right) {\tilde \nabla}{\cal Q} \,,\label{eq:crossprod}
\end{equation}
where  
$
\tilde \nabla=\left({\partial}/{\partial \theta} ,
({1}/{\sin \theta}){\partial}/{\partial \phi} \right)
$.
The companion paper  \citep{biaspaper} shows that the Taylor expansion in the anisotropy for the angular variation, $\cal Q$, of $M_\star$  and  $\dot M_\star$  at fixed distance
$r$ from the saddle scales like
\begin{equation} 
\Delta M_\star\propto \xi_{20}(r) {\cal Q}(\theta,\phi)\,,  \label{eq:defDeltaM} 
\end{equation}
and 
\begin{equation}
  \Delta \dot M_\star \propto
    \bigg[\xi_{20}'(r)- \frac{\sigma-\boldsymbol{\xi'}\cdot \boldsymbol{\xi}}{\sigma^2-\boldsymbol{\xi}\cdot \boldsymbol{\xi}}\xi_{20}(r)\bigg]
 {\cal Q}(\theta,\phi)\,, \label{eq:defDeltaDotM} 
 \end{equation}
in terms of the  variance
\begin{equation}
  \sigma^2(R) =
  \int\dd k \frac{k^2P(k)}{2\pi^2} W^2(kR)\,, \label{eq:sigma2}
\end{equation}
 and the radius dependent vectors 
\begin{align}
  \boldsymbol{\xi}(r) &\equiv
  \{\xi_{00}(r),\sqrt{3}\xi_{11}(r)r/R_\star,\sqrt{5}\xi_{20}(r)\}
\label{eq:xisad}\,, \\
  \boldsymbol{\xi}'\!(r) &\equiv
  \{\xi_{00}'(r),\sqrt{3}\xi_{11}'(r)r/R_\star,\sqrt{5}\xi_{20}'(r)\}\,,
\label{eq:xi'sad}
\end{align}
where 
\begin{equation}
  R^2_\star \equiv
  \int\dd k \frac{P(k)}{2\pi^2} \frac{W^2(k\Rs)}{\sigmas^2}\,,
\label{eq:sadscale}
\end{equation}
with $P(k)$ the underlying power spectrum, $W(k)$ the top-hat filter in Fourier space, $\sigma_\S=\sigma(R_\S)$, while  the  finite separation correlation
functions, $\xi_{\alpha\beta}(r, R, \Rs)$ and $\xi'_{\alpha\beta}(r, R, \Rs)$ are defined as
\begin{align}
\label{eq:xi_xi}
   \xi_{\alpha\beta} &\equiv \int \dd k \frac{k^2P(k)}{2\pi^2}
  W(kR)\frac{W(k\Rs)}{\sigmas}
  \frac{j_\alpha(kr)}{(kr)^\beta}, \\ 
  \xi'_{\alpha\beta} & \equiv
  \int \dd k \frac{k^2P(k)}{2\pi^2} W'\!(kR)
  \frac{W(k\Rs)}{\sigmas}\frac{j_\alpha(kr)}{(kr)^\beta}\,,
  \label{eq:xi_xiprime}
\end{align}
where $j_\alpha(x)$  are the spherical Bessel functions of the first kind and prime denote derivate with respect to $\sigma$. Note that Equation~\eqref{eq:defDeltaDotM}  clearly highlights the shifted variance, $\sigma^2 -\boldsymbol{\xi}\cdot \boldsymbol{\xi}$ which contributes to the difference between $\Delta M_\star$ and $\Delta \dot M_\star$.
From Equation~\eqref{eq:defDeltaDotM}, since the square bracket is not proportional to $\xi_{20}$ as in equation~\eqref{eq:defDeltaM}, it follows that the cross product in Equation~\eqref{eq:crossprod} is non zero, which in turn implies that the contours of mass and accretion rate differ.

\section{Medians of distributions}
\label{sec:appendix_medians}

This Appendix gathers Tables of medians with corresponding error bars used in previous sections.

\begin{table*}
\centering
\begin{threeparttable}
\caption{Medians of $D_{\mathrm{skel}}/\left<D_\mathrm{z}\right>$ for Figure~\ref{Fig:appendix_dskel_dnode} }
\label{tab:mass_gradients_dskel_dnode}
\begin{tabular*}{0.8\textwidth}{@{\extracolsep{\fill}}lccc}
\hline
\hline
\multirow{3}{*}{selection\tnotex{tnote:panels} }&  \multirow{3}{*}{mass bin} &  \multicolumn{2}{c}{median\tnotex{tnote:median} }\\
 & & \multicolumn{2}{c}{$D_{\mathrm{skel}}/\left<D_\mathrm{z}\right>$}  \\
 & & before reshuffling\tnotex{tnote:reshuffling}  & after reshuffling \\
\hline
\hline
\multirow{3}{*}{all galaxies}& $\log \, (\Mstardot)  \geq 11 $ &  0.27 $\pm$ 0.01 &  0.33 $\pm$ 0.02  \\
                                  & $ 11 > \log \, (\Mstardot) \geq 10.7 $& 0.36 $\pm$ 0.01 &  0.37 $\pm$ 0.01  \\
                                  & $ 10.7 > \log \, (\Mstardot) \geq 10.46$ & 0.40 $\pm$ 0.01 &  0.38 $\pm$ 0.01  \\
\hline                                  
\multirow{3}{*}{$d^{\mathrm{node}}_{\mathrm{min}}$ = 3.5 Mpc }& $\log \, (\Mstardot)  \geq 11 $ &  0.38 $\pm$ 0.01 &  0.46 $\pm$ 0.02 \\
                                  & $ 11 > \log \, (\Mstardot) \geq 10.7 $& 0.46 $\pm$ 0.01 &  0.47 $\pm$ 0.01 \\
                                  & $ 10.7 > \log \, (\Mstardot) \geq 10.46$ & 0.51 $\pm$ 0.01 &  0.47 $\pm$ 0.01   \\                                                                    
\hline
\end{tabular*}
\begin{tablenotes}
     \item\label{tnote:panels} panels of Figure~\ref{Fig:appendix_dskel_dnode} 
     \item\label{tnote:median} medians of distributions as indicated in Figure~\ref{Fig:appendix_dskel_dnode} by a vertical lines; 
     errors are computed as in Table~\ref{tab:medians_dskel_dwall}          
      \item\label{tnote:reshuffling} randomisation of $D_{\mathrm{skel}}$ in bins of $D_{\mathrm{node}}$                     
    \end{tablenotes}
\end{threeparttable}
\end{table*}

\begin{table*}
\begin{threeparttable}
\caption{Medians of $D_{\mathrm{wall}}/\left<D_\mathrm{z}\right>$ for Figure~\ref{Fig:appendix_dwall_dskel}}
\label{tab:mass_gradients_dwall_dskel}
\begin{tabular*}{0.8\textwidth}{@{\extracolsep{\fill}}lccc}
\hline
\hline
\multirow{3}{*}{selection\tnotex{tnote:panels} }&  \multirow{3}{*}{mass bin} &  \multicolumn{2}{c}{median\tnotex{tnote:median}} \\
 & &\multicolumn{2}{c}{$D_{\mathrm{wall}}/\left<D_\mathrm{z}\right>$} \\
 & & before reshuffling\tnotex{tnote:reshuffling}   & after reshuffling \\
\hline
\hline
\multirow{3}{*}{$d^{\mathrm{node}}_{\mathrm{min}}$ = 3.5 Mpc}& $\log \, (\Mstardot)  \geq 11 $ &   0.234 $\pm$ 0.005 &  0.258 $\pm$ 0.011 \\
                                  & $ 11 > \log \, (\Mstardot) \geq 10.7 $&  0.279 $\pm$ 0.003 &  0.278 $\pm$ 0.005 \\
                                  & $ 10.7 > \log \, (\Mstardot) \geq 10.46$ &  0.295 $\pm$ 0.003 &  0.292 $\pm$ 0.004\\
\hline                                  
\multirow{3}{*}{$d^{\mathrm{node}}_{\mathrm{min}}$ = 3.5 Mpc, $d^{\mathrm{skel}}_{\mathrm{min}}$ = 2.5 Mpc}& $\log \, (\Mstardot)  \geq 11 $ &  0.334 $\pm$ 0.007 &  0.379 $\pm$ 0.028\\
                                  & $ 11 > \log \, (\Mstardot) \geq 10.7 $&   0.381 $\pm$ 0.004 &  0.386 $\pm$ 0.011\\
                                  & $ 10.7 > \log \, (\Mstardot) \geq 10.46$ &  0.403 $\pm$ 0.004 &  0.398 $\pm$ 0.008  \\                                                                    
\hline
\end{tabular*}
\begin{tablenotes}
     \item\label{tnote:panels} panels of Figure~\ref{Fig:appendix_dwall_dskel} 
     \item\label{tnote:median} medians of distributions as indicated in Figure~\ref{Fig:appendix_dwall_dskel} by a vertical lines; errors are computed as in Table~\ref{tab:medians_dskel_dwall} 
     \item\label{tnote:reshuffling} randomisation of $D_{\mathrm{wall}}$ in bins of $D_{\mathrm{skel}}$                     
    \end{tablenotes}
\end{threeparttable}
\end{table*}


\begin{table*}
\begin{threeparttable}
\caption{Medians for the PDFs displayed in Figure~\ref{Fig:PDF_dskel_dtfe_reshuff_dmatching}: small-scale density}
\label{tab:medians_dskel_dtfe_reshuff_dmatching}
\begin{tabular*}{0.8\textwidth}{@{\extracolsep{\fill}}lccccc}
\hline
\hline
& \multirow{3}{*}{selection\tnotex{tnote:panels}} & \multirow{3}{*}{bin} &  \multicolumn{3}{c}{median\tnotex{tnote:median}} \\
& & &  \multicolumn{3}{c}{$D_{\mathrm{skel}}/\left<D_\mathrm{z}\right>$ } \\
\cline{4-6}
& & & original\tnotex{tnote:before}  & reshuffling\tnotex{tnote:after} & matching\tnotex{tnote:dmatching}\\
\hline
\hline
\multirow{6}{*}{Masses}& \multirow{3}{*}{All galaxies }& $\log \, (\Mstardot)  \geq 11 $ &  0.379 $\pm$ 0.009 &  0.397 $\pm$ 0.009 & 0.378 $\pm$ 0.01\\
                  &                & $ 11 > \log \, (\Mstardot) \geq 10.7 $& 0.456 $\pm$ 0.007 &  0.459 $\pm$ 0.006 & 0.393 $\pm$ 0.009\\
                   &               & $ 10.7 > \log \, (\Mstardot) \geq 10.46$ & 0.505 $\pm$ 0.006 &  0.495 $\pm$ 0.006 & 0.406 $\pm$ 0.008\\
\cline{2-6}
& \multirow{3}{*}{SF galaxies}&$ \log \, (\Mstardot)  \geq 10.8$ & 0.459 $\pm$ 0.012 &  0.489 $\pm$ 0.013 & 0.458 $\pm$ 0.011 \\
 &                                 &$10.8 > \log \, (\Mstardot) \geq 10.3$ & 0.534 $\pm$ 0.007 &  0.541 $\pm$ 0.008 & 0.479 $\pm$ 0.01\\
  &                                &$10.3 > \log \, (\Mstardot) \geq 9.92$ & 0.578 $\pm$ 0.007 & 0.567 $\pm$ 0.007 & 0.494 $\pm$ 0.006 \\
\hline
\multirow{2}{*}{Types} &\multirow{2}{*}{SF vs passive\tnotex{tnote:SF_q_mass}}& star-forming& 0.504 $\pm$ 0.008 & 0.508 $\pm$ 0.007 & 0.495 $\pm$ 0.006\\
 &                                 & passive& 0.462 $\pm$ 0.007 & 0.458 $\pm$ 0.007 & 0.504 $\pm$ 0.006\\                                  
\hline
\end{tabular*}
\begin{tablenotes}
     \item\label{tnote:panels} panels of Figure~\ref{Fig:PDF_dskel_dtfe_reshuff_dmatching}
     \item\label{tnote:median} medians of distributions as indicated in Figure~\ref{Fig:PDF_dskel_dtfe_reshuff_dmatching} by a vertical lines; errors are computed as in Table~\ref{tab:medians_dskel_dwall} 
     \item\label{tnote:before} as in Table~\ref{tab:medians_dskel_dwall} for $D_{\mathrm{skel}}/\left<D_\mathrm{z}\right>$                                     
     \item\label{tnote:after} reshuffling is done in bins of DTFE density (see main text for more details)
     \item\label{tnote:dmatching} medians for the density-matched sample, where the density considered is DTFE
     \item\label{tnote:SF_q_mass} only galaxies with stellar masses  $\log \, (\Mstardot) \geq 10.46$ are considered
    \end{tablenotes}
\end{threeparttable}
\end{table*}


\begin{table*}
\begin{threeparttable}
\caption{Medians for the PDFs displayed in Figure~\ref{Fig:PDF_dskel_reshuff_HzAGN}}
\label{tab:medians_dskel_dens_reshuff_hagn}
\begin{tabular*}{0.8\textwidth}{@{\extracolsep{\fill}}lcccc}
\hline
\hline
\multirow{4}{*}{selection\tnotex{tnote:panels}} & \multirow{4}{*}{bin} &  \multicolumn{3}{c}{median\tnotex{tnote:median}} \\
& &  \multicolumn{3}{c}{$D_{\mathrm{skel}}$ [Mpc] } \\
\cline{3-5}
& & \multirow{2}{*}{original\tnotex{tnote:before}}  & \multicolumn{2}{c}{after reshuffling\tnotex{tnote:after}}\\
& & & DTFE & G5Mpc\\
\hline
\hline
 \multirow{3}{*}{Mass }& $\log \, (\Mstardot)  \geq 10.8 $ &  1.34 $\pm$ 0.09 &  1.26 $\pm$ 0.08 & 1.72 $\pm$ 0.1\\
                                 & $ 10.8 > \log \, (\Mstardot) \geq 10.4 $& 1.73 $\pm$ 0.08 &  1.71 $\pm$ 0.06 & 1.82 $\pm$ 0.06\\
                                 & $ 10.4 > \log \, (\Mstardot) \geq 10$ & 1.97 $\pm$ 0.04 &  2.0 $\pm$ 0.05 & 1.86 $\pm$ 0.04\\
\hline	
\multirow{3}{*}{\ssfr}&$ -10.8 > \log \, (sSFR/\mathrm{yr})$ & 1.46 $\pm$ 0.07 & 1.61  $\pm$ 0.07 & 1.74 $\pm$ 0.08\\
                                 &$-10.4 > \log \, (sSFR/\mathrm{yr}) \geq -10.8$ & 1.88 $\pm$ 0.06 &  1.89 $\pm$ 0.06 & 1.81 $\pm$ 0.06 \\
                                 &$\log \, (sSFR/\mathrm{yr}) \geq -10.4$ & 2.0 $\pm$ 0.04 & 1.9 $\pm$ 0.05 & 1.91 $\pm$ 0.06\\
\hline
\end{tabular*}
\begin{tablenotes}
     \item\label{tnote:panels} panels of Figure~\ref{Fig:PDF_dskel_reshuff_HzAGN}
     \item\label{tnote:median} medians of distributions as indicated in Figure~\ref{Fig:PDF_dskel_reshuff_HzAGN} by a vertical lines; errors are computed as in Table~\ref{tab:medians_dskel_dwall} 
     \item\label{tnote:before} as in Table~\ref{tab:medians_dskel_dwall_HzAGN} for $D_{\mathrm{skel}}$  (corresponding to the solid lines in Figure~\ref{Fig:PDF_dskel_reshuff_HzAGN})                                    
     \item\label{tnote:after} reshuffling is done in the bins of the DTFE density and the density computed at the scale of 5 Mpc (corresponding to the dashed lines in Figures a and b, respectively)
    \end{tablenotes}
\end{threeparttable}
\end{table*}

\label{lastpage}

\end{document}